\documentclass[11pt,a4paper]{article}
\usepackage{amsfonts}
\usepackage{amssymb}
\usepackage{jheppub}
\usepackage{ulem}
\usepackage{multirow}
\usepackage{float}

\usepackage{cancel}

\usepackage{inputenc}
\usepackage{xspace}
\usepackage{url}

\usepackage{tikz}
\usepackage{pgflibraryarrows}
\usepackage{pgflibrarysnakes}

\newcommand{\bea}{\begin{eqnarray}}
\newcommand{\eea}{\end{eqnarray}}
\newcommand{\bean}{\begin{eqnarray*}}
\newcommand{\eean}{\end{eqnarray*}}
\newcommand{\nn}{\nonumber\\}
\newcommand{\Sl}{\sum\limits}
\newcommand{\blue}{\color{blue}}
\newcommand{\red}{\color{red}}
\newcommand{\green}{\color{green}}

\def\W #1{\widetilde{#1}}

\def\Label#1{\label{#1}%
  \smash{\hbox to0pt{\raise1ex\hbox{\tiny[#1]}\hss}}}
\def\Label#1{\label{#1}}
\renewcommand{\eqref}[1]{eq.~(\ref{#1})}

\newcommand{\figref}[1]{figure~\ref{#1}}

\newcommand{\secref}[1]{section~\ref{#1}}
\newcommand{\appref}[1]{appendix~\ref{#1}}

\renewcommand{\red}{\color{black}}
\renewcommand{\blue}{\color{black}}
\renewcommand{\green}{\color{black}}

\allowdisplaybreaks
\title{Leading order multi-soft behaviors of tree amplitudes in NLSM}
\author[a]{Yi-Jian Du}

\emailAdd{yijian.du@whu.edu.cn}
\author[b]{Hui Luo}

\affiliation[a]{Center for Theoretical Physics,
School of Physics and Technology,
Wuhan University,\\
299 Bayi Road, Wuhan 430072,
China}
\affiliation[b]{II. Institut f\"{u}r Theoretische Physik, Universit\"{a}t Hamburg, \\Luruper Chaussee 149, D- 22761 Hamburg, Germany}

\emailAdd{hui.luo@desy.de}
\abstract{{In this paper}, we investigate multi-soft behaviors of tree amplitudes in nonlinear sigma model (NLSM). 
The leading behaviors of amplitudes with odd number of all-adjacent soft pions are zero.
 We further propose and prove {\blue that leading soft factors of amplitudes with even number all-adjacent soft pions can be expressed} in terms of products of the leading order Berends-Giele sub-currents in Cayley parametrization. 
 Each sub-current in the expression contains at most one hard pion. 
Discussions are generalized to amplitudes containing arbitrary number of nonadjacent soft blocks:  
{The leading behaviors of amplitudes where at least one soft block has odd number of adjacent soft pions are zero;}  
{The leading soft factors for amplitudes where all soft blocks containing even number of soft pions  are given by products of soft factors for {\blue these} blocks.}  }
\keywords{Scattering Amplitudes, Nonlinear Sigma Model, Soft Limit}
\begin{document}
\maketitle
\section{Introduction}
Symmetry is one of the most critical topic in physics.
{\green Refer to the global symmetry, which is alway broken to describe the real world, it can be broken spontaneously by gaining certain vacuum expectation values or explicitly by terms violating the symmetry.}
In particular, once a global symmetry $G$ of a theory is spontaneously broken into its subgroup $H$, massless Goldstone bosons are in one-to-one correspondence to the broken generators.
The unbroken symmetry generators $T_i$ and the broken generators $X_a$ {satisfy the following} schematic commutation relations
\bea
[T, T]\sim T, \, [X, T]\sim X, \, [X,X]\sim T.
\eea
This global symmetry $G$ can be realized by properly defined fields of Goldstone bosons, whose behaviors are able to be described by the nonlinear sigma model (NLSM)\cite{Coleman:1969sm,Callan:1969sn}.
Because the scattering amplitudes at any point of the vacuum moduli {space} should be identical, the vacuum structure after global symmetry spontaneously breaking can be probed by certain structures of scattering amplitudes \cite{ArkaniHamed:2008gz}. 

{To reveal states of Hilbert space} in different vacua through scattering amplitudes, {\green one {should} study the behaviors of amplitudes with additional Goldstone bosons of zero momentum by regulating these constant scalar fields with tiny momenta and sending them to zero eventually with a very careful analysis.}
More precisely speaking, if a theory after symmetry breaking has a continuous vacuum, one vacuum can be arrived by expanding around another one
\bea\label{vacuum-expand}
|\Omega_\theta\rangle
=|\Omega \rangle+|\Omega^{(1)}\rangle+|\Omega^{(2)}\rangle+\dots,
\eea
where the variation $|\Omega^{(n)}\rangle$ should contain $n$ soft pions\footnote{pion is a physical synonym for Goldstone boson. We don't distinguish them in this paper.} information.
This implies {that} the vacuum variations might be discovered by studying soft limits of amplitudes.
For example, the first order variation can be understood by taking a single soft pion, which vanishes eventually and is well known as "Adler zero" \cite{Adler:1964um,Susskind:1970gf}.
The second order variation can be described by double soft pions emissions, which is related to an $[X^\alpha, X^\beta]$ transformation on the amplitude of hard pions.
Moreover, the double soft limits are studied up to the next leading order recently \cite{ArkaniHamed:2008gz, Cachazo:2015ksa, Du:2015esa,Low:2015ogb,He:2016vfi}\footnote{Discussions on soft limits in theories with conformal symmetry breaking and supersymmetry breaking can be found in \cite{Boels:2015pta, Huang:2015sla, DiVecchia:2015jaq, Chen:2014xoa,Kallosh:2016qvo,Kallosh:2016lwj}.}.  
Along this line, if studying higher order variations, $|\Omega^{(3)}\rangle$, $|\Omega^{(4)}\rangle$ and so on, we need to understand multi-soft behaviors of amplitudes.



 In the present work, we focus on the non-linear sigma model (NLSM) from spontaneously global symmetry breaking  $SU(N)\times SU(N)\rightarrow SU(N)$.
{Analogous to the color decomposition in Yang-Mills theory, the flavor structure in NLSM can be  separated from kinematic factors which are involved in flavor-ordered partial amplitudes. 
Once we know all multi-soft behaviors for partial amplitudes, we know the behaviors for full amplitudes.}
To study the multi-soft behaviors of partial amplitudes, {we could} either take soft limits consecutively or simultaneously.
{Although there might be symmetry insights for both consecutive and simultaneous soft limits 
 (see e.g, \cite{McLoughlin:2016uwa, Lipstein:2015rxa}),  one can obtain the consecutive soft behaviors by applying simultaneous ones in principle.}
{In this paper, we systematically study all leading order simultaneous soft behaviors of tree-level partial amplitudes $A(1,2,\cdots,2n)$ in NLSM with arbitrary number
 of soft pions. 
 Through denoting the momenta of all soft pions $i\in\{S\}$ by $k_i=\tau q_i$, the partial amplitude becomes a function of $\tau$, i.e., $A(\tau)$. 
 Thus the soft behavior is described by $A(\tau)$ with $\tau\to 0$. }
The possible leading order of $A(\tau)$ in NLSM is the coefficient of $\tau^0$ order in Taylor's expansion.

{This work is devoted to the study of all the $\tau^0$ behaviors of partial amplitudes in NLSM. We first study the behaviors of amplitudes with all-adjacent soft pions. Such amplitudes are classified according to whether the number of soft pions is odd or even. The behaviors are given as follows:
\newline
{\textit{a)} \textit{The $\tau^0$ order of amplitude $A\bigl(1\cdots, a, \W {a+1},\cdots,\W {a+2i+1}, a+2i+2,\cdots,2n\bigr)$, which contains odd number of all-adjacent soft pions {\blue $\W{a+1},\cdots, \W{a+2i+1}$}, has to vanish.}} 
This result will be illustrated via on-shell recursion \cite{Britto:2004ap,Britto:2005fq} as well as off-shell recursion \cite{Kleiss:1988ne} in NLSM (see \cite{Kampf:2012fn,Kampf:2013vha,Cheung:2015ota,Luo:2015tat}). 
When $i=0$, where only one soft pion is contained, the vanishing behavior agrees with Alder's zero \cite{Adler:1964um}.
\newline
{\textit{b)} \textit{The $\tau^0$ order of amplitude $A\bigl(1,\cdots,a, \W{a+1},\cdots, \W{a+2i}, a+2i+1,\cdots, 2n\bigr)$, which contains even number of all-adjacent soft pions {\blue $\W{a+1},\cdots, \W{a+2i}$} and at least four hard pions \footnote{{\red The special case with only two hard pions, no matter they are adjacent or not, should have vanishing $\tau^0$ order.}},  is given by}
\bea
&&A\bigl(1,\cdots,a, \W{a+1},\cdots, \W{a+2i}, a+2i+1,\cdots, 2n\bigr)\nn
&=&\mathbb{S}^{(0)}_{a,a+2i+1}A\bigl(1,\cdots, a,a+2i+1,\cdots,2n\bigr)+\mathcal{O}(\tau),\Label{eq:EvenAdjSoft}
\eea}
where the soft factor $\mathbb{S}^{(0)}_{a,a+2i+1}$ is defined as
\bea
\mathbb{S}^{(0)}_{a,a+2i+1}=\Sl_{D\in\text{divisions}\{a,\W {a+1},\cdots,\W {a+2i}, a+2i+1\}} \left(-{1\over 2F^2}\right)^{N(D)-2 \over 2}\prod\limits_{j=1}^{N(D)}J^{(0)}\left(D_j\right).\Label{eq:EvenAdjSoftFactor}
\eea
Here, all possible divisions $D$ of the ordered set $\{a,\W {a+1},\cdots,\W {a+2i}, a+2i+1\}$ such that each subset contains odd number of elements are summed over. Coupling constant is $-{1\over 2F^2}$ and $N(D)$ is the number of subset $\{D_j\}$ for a given division $D$.
The number of subsets should be more than one.
For example, while setting $a=1$ and considering the behavior of amplitudes with four soft pions, we have the following divisions {\blue of $\{1, \W 2,\W 3,\W 4,\W 5, 6\}$}
\bea
&&\bigl\{\{1,\W2,\W3,\W4,\W5\},\{6\}\bigr\},~~~~~~~~~~~~\bigl\{\{1,\W 2,\W 3\},\{\W 4,\W5, 6\}\bigr\},~~~~~~~~~~~~~\bigl\{\{1\},\{\W 2,\W 3,\W4,\W5, 6\}\bigr\},\nn
&&\bigl\{\{1,\W2,\W3\},\{\W4\},\{\W5\},\{6\}\bigr\},~~~~~~\bigl\{\{1\},\{\W 2,\W 3,\W 4\},\{\W5\}, \{6\}\}\bigr\}, ~~~~~\bigl\{\{1\},\{\W 2\},\{\W 3,\W 4,\W5\}, \{6\}\bigr\},\nn
&&\bigl\{\{1\},\{\W 2\},\{\W 3\},\{\W 4,\W5, 6\}\bigr\},~~~~~~\bigl\{\{1\},\{\W2\},\{\W3\},\{\W4\},\{\W5\},\{6\}\bigr\}.\Label{eq:4SoftDivisions}
\eea
In \eqref{eq:EvenAdjSoftFactor}, $J^{(0)}({D}_j)$ denotes the $\tau^0$ order of Berends-Giele currents $J({D}_j)$ of $U(N)$ NLSM (The on-shell amplitudes of $U(N)$ and $SU(N)$ NLSM are equivalent) in Cayley parametrization \cite{Kampf:2013vha}. From the definition of divisions, we know that the even number all-adjacent soft behavior is expressed by the leading orders of two kinds of Berends-Giele sub-currents: \textit{i)} sub-currents containing only soft external pions, e.g., $J^{(0)}(\W 2,\W 3,\W 4)$ from the division $\bigl\{\{1\},\{\W 2,\W 3,\W 4\},\{\W5\}, \{6\}\}\bigr\}$, \textit{ii)} sub-currents where only the first or last pion is hard and all others are soft, e.g.,  $J^{(0)}(1,\W2,\W3)$ from the division $\bigl\{\{1,\W2,\W3\},\{\W4\},\{\W5\},\{6\}\bigr\}$. {\red Actually, we will argue that these two kinds of sub-currents only have $\tau^0$ order term, i.e., $J^{(0)}({D}_j)=J({D}_j)$. Nevertheless, we always use $J^{(0)}({D}_j)$ in this paper to emphasize that there is no $\tau$ dependence in the final expression \eqref{eq:EvenAdjSoftFactor}.} The behavior \eqref{eq:EvenAdjSoftFactor} will be proved by Berends-Giele recursion.

We further generalize the off-shell recursion proofs of behaviors \textit{a}) and \textit{b)}
to amplitudes containing more than one soft blocks and obtain the following results:
\newline
{\textit{c)} \textit{The $\tau^0$ term of an amplitude containing nonadjacent soft blocks, if one of which is of odd number soft pions, vanishes.}}
\newline
{\textit{d)} \textit{The $\tau^0$ term of an amplitude with {\red at least four hard pions and} nonadjacent soft blocks, where all blocks contain even number of soft pions, is a product of soft factors $\mathbb{S}^{(0)}$ for each block.}}}

This paper is organized as follows. In \secref{Sec:Review}, we review off-shell and on-shell recursion relations for tree amplitudes in NLSM.
In \secref{Sec:OddSoft}, we illustrate that the leading order of partial amplitudes with adjacent soft pions of odd number is always vanishing.
Behaviors of partial amplitudes with even number adjacent soft pions are studied \secref{Sec:EvenSoft}.
In \secref{sec:NonadjacetBlocks}, we generalize our discussions to amplitudes with nonadjacent soft blocks.
In \secref{Sec:Conclusion}, we conclude this work and discuss the physical insight of the multi-soft pions.
We leave a tedious proof for {\blue a case of} even number soft pion in the appendix as well as an explicit result of six soft pions. 

\section{A review of off-shell and on-shell recursions in NLSM}\label{Sec:Review}
{In this section, we  briefly review both the off-shell Berends-Giele recursion in Cayley parameterization and on-shell recursion in NLSM, which will be used to prove multi-soft behaviors in this paper.}

\subsection{Feynman rules and Berends-Giele recursion relations}

{The on-shell amplitudes for $SU(N)$ NLSM are shown to be same with those in $U(N)$ NLSM, thus we can just consider amplitudes in $U(N)$ NLSM. } Tree amplitudes of $U(N)$ NLSM can be decomposed into group structure factors and kinematic part {as follows}
\bea
M(1^{a_1},\dots,n^{a_n})=\Sl_{\sigma\in S_{n-1}}\mathrm{Tr}(T^{a_{1}}T^{a_{\sigma_2}}\dots T^{a_{\sigma_n}})A(1,\sigma).\Label{eq:Trace form}
\eea
Due to the analogy between the group structure in NLSM and the color factors in Yang-Mills theory, in the following discussion, the group structure factors in NLSM are {mentioned} as  factors as well, and the kinematic part $A(1,\sigma)$ are color-ordered amplitudes which are evaluated by Feynman rules in Cayley parametrization, i.e.
{\bea
V_{2n+1}=0,\qquad V_{2n+2}=\left(-{1\over 2F^2}\right)^n\left(\Sl_{i=0}^nk_{2i+1}\right)^2=\left(-{1\over 2F^2}\right)^n\left(\Sl_{i=0}^nk_{2i+2}\right)^2.\Label{eq:Feyn-rules}
\eea
Here, $k_j$ is the momentum of the external particle $j$ and note that the momentum conservation should be preserved.}
Apparently, in Cayley parametrization, only the scattering amplitudes {with even number of external pions} are nonzero, which of course is a fact independent of field definitions.
Thus in this paper, we always consider pion amplitudes of even number {external lines}.

With Feynman rules \eqref{eq:Feyn-rules} at hand, tree-level currents can be constructed through Berends-Giele recursion relations ({\blue see \cite{Kampf:2012fn,Kampf:2013vha}})
\bea
&&J(2,...,2n)\Label{eq:B-G}\\
&=&\frac{i}{P_{2,2n}^2}\Sl_{m=2}^n\Sl_{\text{Divisions}}i V_{2m}(k_1=-P_{2,2n},P_{A_1},\cdots,P_{A_{2m-1}})\times\prod\limits_{k=1}^{2m-1} J(A_{k}),\nonumber
\eea
where {$k_1=-P_{2,n}\equiv-(k_2+k_3+\dots+k_n)$} is the momentum of the off-shell pion ``$1$".
In the second sum,  ``Divisions" denotes all possible partitions of on-shell particles $\{2,\dots,2n\}\to \{A_1\},\dots,\{A_{2m-1}\}$ with odd number elements in each subset $\{A_i\}$.
The starting point of this off-shell recursion is $J(2)=J(3)=\dots=J(n)=1$.
{While multiplied by  $-k_1^2\to 0$,
 the current (see \eqref{eq:B-G}) becomes on-shell amplitude $A(1,\cdots,n)$.}

\subsection{On-shell recursion relations of effective theories}\label{sec:OnShellRecursion}
In \cite{Cheung:2015ota}, on-shell recursion relations are proposed for effective theories of scalars.
In particular, those amplitudes with vanishing single soft limits can be calculated by employing new rescaling momentum shifts, which works in NLSM, Dirac-Born-Infeld and Galileon theories. {Now let us review the on-shell recursion for NLSM.}

{For $2n$-point partial amplitude $A(1,\cdots,2n)$  in NLSM, we deform the momenta of all external pions as}
\bea
 k_{\hat{i}} = (1-a_i z) \, k_i \,~~~~~~~(1\leq i\leq 2n).
\eea
The $2n$-point partial amplitude $A_n(1,\cdots,2n)$ can be constructed from lower-point ones
\bea
A(1,\cdots,2n) = \sum_I {1 \over P_I^2} {A_L(z^-_I)A_R(z^-_I) \over \Bigl(1-z^-_I/z^+_I\Bigr) {F_n(z^{-}_I)}} + (z^-_I \leftrightarrow z^+_I) \, ,\Label{eq:On-shellRecursion}
\eea
 where all possible factorized channels $I$ are summed and the order of external pions are preserved.
 Given a partition $I$, ${P_I = \sum_{i \in I} k_i}$ and $z^{\pm}_I$ are the solutions to the on-shell condition of the internal propagator
\bea
 0=\Bigl(P_I - z \sum_{i \in I} a_i k_i\Bigr)^2 =P^2_I - 2z \Bigl(\sum_{i \in I} a_i k_i\Bigr) \cdot P_I +z^2 \Bigl(\sum_{i \in I} a_i k_i\Bigr)^2  \, ,
\eea
which results in the solutions
\bea
 z^{\pm}_I = { \big(\sum_{i \in I} a_i k_i\big) \cdot P_I  \pm \sqrt{\big[ (\sum_{i \in I} a_i k_i) \cdot P_I\big]^2 - P^2_I \, \big(\sum_{i \in I} a_i k_i\big)^2 } \over \big(\sum_{i \in I} a_i k_i\big)^2 } \,.\Label{eq:solution}
\eea
The function ${F_n(z)}$ in \eqref{eq:On-shellRecursion} is defined by
\bea
F_n(z)=\prod_{i=1}^n (1-a_i z)\,.
\eea

\section{Amplitudes with  adjacent soft pions of odd number}\label{Sec:OddSoft}

The {vanishing} of $\tau^0$ order with a single soft pion is already well known as Adler's zero.
In this section, we generalize the Adler's zero to amplitudes with odd number {of all-adjacent} soft pions.
Using both on-shell and off-shell recursion relations, we illustrate that the $\tau^0$ order of color-ordered tree amplitudes which containing arbitrary odd number of all-adjacent soft pions has to vanish.


\subsection{On-shell recursion approach}\label{on-shell-odd}
Now let us study amplitudes with odd number of all-adjacent soft pions whose momenta are denoted by $\tau k_{i\in \{S\}}$, using the  on-shell recursion reviewed in \secref{sec:OnShellRecursion}.

First of all, {the starting point of  on-shell recursion approach is that \textit{four-point amplitude with one or three soft pions should  vanish at $\tau^0$ order,} which can be checked from direct evaluation by the Feyman rules \eqref{eq:Feyn-rules}.}
One may notice that, according to momentum conservation, {when three pions in a four-point amplitude are soft, the fourth one have to be soft too.}
If {the momenta of all pions are of $\tau$ order}, it seems meaningless to discuss soft limits any more.
However, we are not really discussing about $2n-1$ soft limits of $2n$-point amplitude $A(1,\dots, 2n)$, and this special case only plays a role in the subamplitudes while applying on-shell recursion relations.
Thus, the maximal odd number which makes the soft limits of $A(1,\dots, 2n)$ interesting and meaningful is $2n-3$.

{Assuming that the $\tau^0$ order of all $2m~(2m<2n)$-point amplitudes with odd number of soft pions vanish, we now use the on-shell recursion \eqref{eq:On-shellRecursion} to investigate the behavior of $2n$-point amplitude. In the on-shell recursion expression  \eqref{eq:On-shellRecursion} of $A(1,\dots, 2n)$,} subamplitudes $A_L$ and $A_R$ contain at least four on-shell pions respectively: one comes from taking the internal line on-shell and others are external pions.
According to different distributions of the $2i+1$ soft pions, we classify terms in the recursion expression  \eqref{eq:On-shellRecursion} as follows.
\begin{itemize}
\item[(a)]  One subamplitude contains only soft external pions, while the other one contains hard external and probably soft pions.
\item[(b)]  One subamplitude contains only hard external pions, while the other one contains both soft and at least two hard external pions.
\item[(c)]  Both subamplitudes contain hard and soft pions.
\end{itemize}
{In all three cases, the solution (\ref{eq:solution}) can only start from $\tau^0$. This is because the leading behaviors of denominator and numerator of \eqref{eq:solution} should be in a same order. Thus we only need to discuss the behaviors of subamplitudes $A_L$, $A_R$ and propagator ${1\over P_I^2}$. }

\textit{For the terms in category (a)}, without loss of generality, consider the case where $A_L$ contains only soft external pions, while $A_R$ contains both soft and hard external pions.
Since all external pions in $A_L$ are soft, each of which contributes a factor $\tau$ in $A_L$ from its momentum. It pays to discuss the dimension and scale analysis about the amplitudes.
In the NLSM with two derivatives, no matter what kind of parameterization is implemented, the vertex is of $p^2$ mass dimension order, the same as propagators.
At classical level, the number of vertices $N_{\text{vertices}}$ and the number of propagators $N_{\text{propagators}}$ obeys $N_{\text{vertices}}=N_{\text{propagators}}+1$.
Thus mass dimension of amplitudes should be $p^2$ order compensated by decay constant $F^2$.
All soft external legs in $A_L$ should make {the amplitude} of $\tau^2$ {order}.
The ${1\over P_I^2}$ in \eqref{eq:On-shellRecursion} contributes a ${1\over \tau^2}$.
In $A_R$, there is at least one soft pion, which comes from the internal propagator, and probably other even number soft external pions.
In total, $A_R$ contains odd number of soft pions.
According to the inductive assumption, the leading order of $A_R$ in soft limit is $\mathcal{O}(\tau)$.
All together, terms of in category (a) obey a soft behavior of $\mathcal{O}(\tau)$.

\textit{For the terms in category (b)}, the subamplitude $A_L$ containing only hard pions and ${1\over P^2_I}$ are both of the order $\tau^0$, while the subamplitude $A_R$ contains odd number of soft pions thus should be of order $\mathcal{O}(\tau)$ from the inductive assumption.

\textit{For the terms in category (c)}, assume $A_L$ contains soft pions of odd number {while} $A_R$ contains soft pions of even number.
{The subamplitudes $A_R$ with even number all-adjacent soft pions, which will be systematically discussed in the next section, start from $\tau^0$ order. Remembering that the propagator ${1\over P^2_I}$ also starts from  $\tau^0$ order and the $\tau^0$ order of $A_L$ with adjacent odd number soft pions should vanish, we conclude that terms of this category must be zero.  }
\textit{A very special case in this category} is that $A_R$  contains only one hard external pion. In this case, the ${1\over P^2_I}$ starts from $\mathcal{O}(\tau^{-1})$, $A_R$ is a subamplitude with soft pions of even number and two hard legs (one is the internal propagator), which behaves as $\mathcal{O}(\tau)$ according to {\red discussion in the next section.} {The other subamplitude $A_L$,} containing soft pions of odd number, behaves as $\mathcal{O}(\tau)$. Thus, {terms of this category also} behaves as $\mathcal{O}(\tau^{1})$ as we expect.

Putting (a), (b) and (c) together, the {vanishing} of $\tau^0$ order of amplitudes in odd-point soft limit is proved by on-shell recursions.

\subsection{Berends-Giele recursion approach}\label{off-shell-odd}
%
\begin{figure}[h!]
  \centering
 \includegraphics[width=0.9\textwidth]{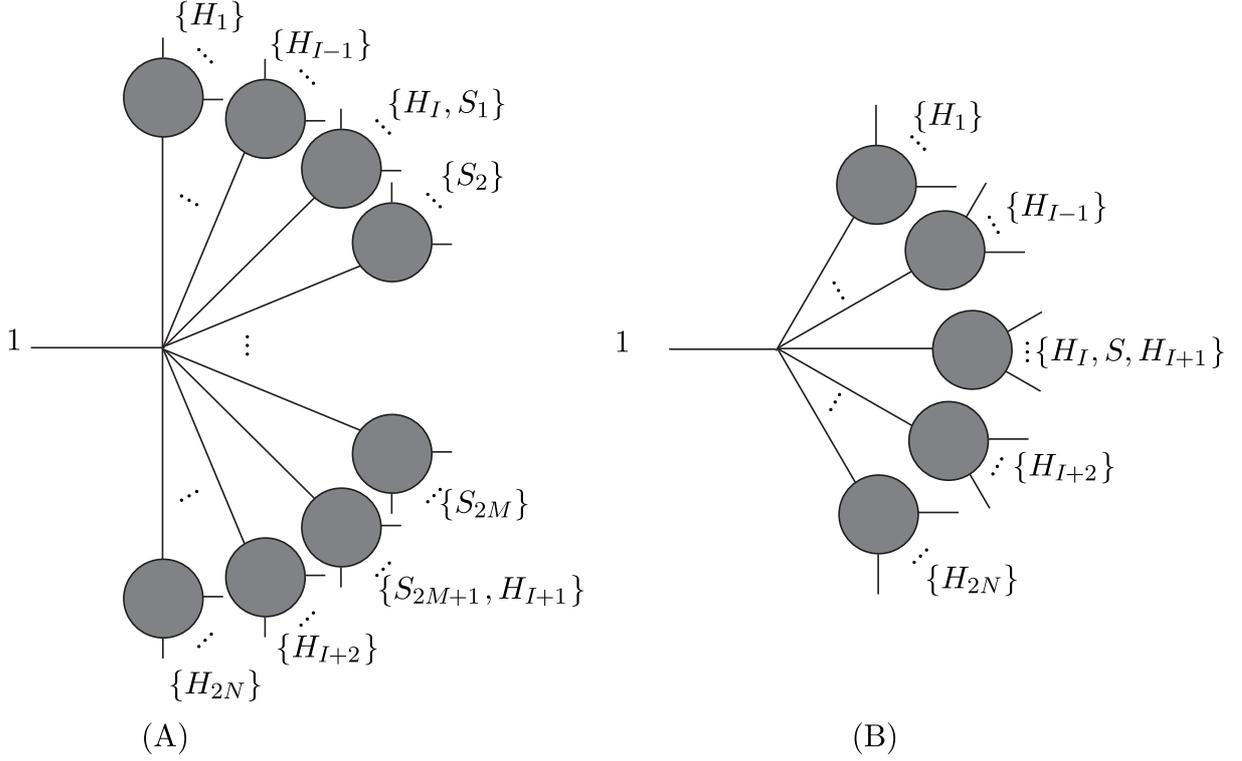}
 \caption{Typical diagrams for currents with odd number soft pions. } \label{OddNumberSoft}
\end{figure}

As a cross check for the {behavior of amplitudes with adjacent odd number soft pions} from on-shell recursion and {as a warm-up for discussions on the behavior with even number soft pions}, we apply the Berends-Giele {recursion} {to prove the vanishing of the $\tau^0$ order of amplitudes containing adjacent odd number soft pions.}

Consider a Berends-Giele current $J(2,\cdots,2n)$ defined by \eqref{eq:B-G} in Cayley parametrization.
If all on-shell pions in this current are soft, namely $J(\W 2,\W 3,\dots, \W{2n})$,
from Feynman rules in \eqref{eq:Feyn-rules}, each vertex in this current contributes a $\tau^2$, while each propagator in \eqref{eq:B-G} contributes a $\tau^{-2}$.
In a tree-level Berends-Giele current, the number of propagators and of vertices are the same.
Thus the behavior of such current is of order $\tau^0$.
Actually, after dividing out all the $\tau$ factors, the $\tau^0$ term {of $J(\W 2,\W 3,\dots, \W{2n})$} is just a current {$J(2,3,\dots, {2n})$} with replacing the momenta $k_1$, ..., $k_{2n}$ by $q_1$, ..., $q_{2n}$.
If there exit more than one hard pions, the Berends-Giele currents with odd number of soft pions are of the generic form
$J\bigl(2\cdots, a, \W {a+1},\cdots,\W {a+2i+1}, a+2i+2,\cdots,2n\bigr)$.
According to {whether $a$ is odd or even},  these currents can be classified into two types.
\begin{itemize}
\item \textit{Type-1:} If $a=2l-1$, {there are even number of  hard pions appear before/after the first/last soft pion.} {The Berends-Giele current of this type behaves as}
    \bea
    J(2,\dots,2l-1,\W{2l},\dots,\W{2l+2i},2l+2i+1,\dots,2n)=0+\mathcal{O}(\tau^1).\Label{Eq:Type-3a}
    \eea
While $l=1$, the current becomes $J(\W{2},\dots,\W{2+2i},2l+2i+1,\dots,2n)$ where the first soft pion is adjacent to the off-shell leg (see \cite{Du:2015esa} for discussion about $l=1$, $i=0$).

\item \textit{Type-2:} If $a=2l$, {there are odd number of hard pions appear before/after the first/last soft pions. The Berends-Giele current of this type behaves as }
    \bea
    &&J(2,\dots,2l,\W{2l+1},\dots,\W{2l+2i+1},2l+2i+2,\dots,2n)\nn
    &=&-\Sl_{D}\prod_{i=1}^{N_D}\left(-{1\over 2F^2}\right)^{{N_D-1\over 2}} J^{(0)}(D_i)+\mathcal{O}(\tau^1).\Label{Eq:Type-3b}
    \eea
    Here, {we summed over all possible divisions $D$ of $2,\dots,2l,\W{2l+1},\dots,\W{2l+2i+1},2l+2i+2,\dots,2n$ such that all the hard pions} $\{2,\dots,2l\}$ and $\{2l+2i+2,\dots,2n\}$ belong to the first and the last set respectively.
    $J^{(0)}(D_i)$ stands for the $\tau^0$ order of current $J(D_i)$.
\end{itemize}
The on-shell limit of the soft behavior \eqref{Eq:Type-3a} apparently gives zero result for $\tau^0$ order.  When we multiply $k_1^2\to 0$ to the second type of soft behavior \eqref{Eq:Type-3b}, we can also see the vanish of $\tau^0$ order of on-shell amplitudes. Thus \eqref{Eq:Type-3a} and \eqref{Eq:Type-3b} precisely produce the expected $\tau^0$ behavior of amplitudes with odd number of all-adjacent soft pions.
{Assuming that all type-1/type-2 currents with fewer external pions satisfy \eqref{Eq:Type-3a}/\eqref{Eq:Type-3b},} let us prove the behaviors (\ref{Eq:Type-3a}) and (\ref{Eq:Type-3b}) {by induction}.

\subsubsection*{Proof of \eqref{Eq:Type-3a}}\label{subsec:Proof(3.1)}

  When we expand $J\bigl(2\cdots, a, \W {a+1},\cdots,\W {a+2i+1}, a+2i+2,\cdots,2n\bigr)$ through Berends-Giele recursions, corresponding diagrams can be further classified to two types: \textit{i)} soft pions living in more than one sub-current, see \figref{OddNumberSoft}. (A); \textit{ii)} all soft pions in a same sub-current, see \figref{OddNumberSoft}. (B).

{Eq. (\ref{Eq:Type-3a}) gives the behavior when $a=2l-1$.}
If there are soft pions of odd number in {the ordered set} $\{H_I, {S_{1}}\}$
\footnote{Here we use $\{S\}$, $\{H\}$ to denote ordered sets of soft and hard pions respectively, which are used throughout this paper.}
(or $\{{S_{2M+1}},H_{I+1}\}$) in \figref{OddNumberSoft}. (A), {the current $J(H_I, {S_{1}})$ (or $J({S_{2M+1}},H_{I+1})$) must be type-1 currents. Thus its $\tau^0$ order must vanish according to the inductive assumption.}


If both sets $\{H_I, {S_{1}}\}$ and $\{{S_{2M+1}},H_{I+1}\}$ in \figref{OddNumberSoft}. (A)  contain soft pions of even number (and hard pions of odd number),  $I$ should be even, i.e., $I=2L$.
The $\tau^0$ order of \figref{OddNumberSoft}. (A) is
\bea
&&i\left(-{1\over 2F^2}\right)^{N+M-1}{i\left(\Sl_{i=1}^{L}P_{H_{2i-1}}+\Sl_{j=L+1}^{N}P_{H_{2j}}\right)^2\over \left(\Sl_{i=1}^{2L}P_{H_i}+\Sl_{i=2L+1}^{2N}P_{H_i}\right)^2}\Label{Eq:Type3I}\\
&\times&\prod_{i=1}^{2L-1}J(H_i)J^{(0)}(H_{2L},{S_1} )\prod_{j=2}^{2M}J^{(0)}({S_j} )J^{(0)}({S_{2M+1}},H_{2L+1})\prod_{l=2L+1}^{2N}J(H_l)+\mathcal{O}(\tau^1).\nonumber
\eea
%

{On the other hand, we turn to \figref{OddNumberSoft}. (B). Since only the currents with odd number of external pions are nonzero and we have odd number soft pions, the sets $\{H_I\}$ and $\{H_{I+1}\}$ in $J(H_I, S,H_{I+1})$ should simultaneously contain odd (or even) number of hard pions. If both $\{H_I\}$ and $\{H_{I+1}\}$ contain odd number of hard pions, the sub-current $J(H_I, S, H_{I+1})$ have to vanish due to the inductive assumption.  Therefore, $\{H_I\}$ and $\{H_{I+1}\}$ should both contain even number of hard pions.
As a result, $I$ should be even, i.e., $I=2L$ for some $L$.}
For a given \figref{OddNumberSoft}. (A), there is always a corresponding \figref{OddNumberSoft}. (B), whose leading order is {following}
\bea
&&i\left(-{1\over 2F^2}\right)^{N-1}{i\left(\Sl_{i=1}^{L}P_{H_{2i-1}}+\Sl_{j=L+1}^{N}P_{H_{2j}}\right)^2\over \left(\Sl_{i=1}^{2L}P_{H_i}+\Sl_{i=2L+1}^{2N}P_{H_i}\right)^2}
\prod_{i=1}^{2L-1} J(H_i)\prod_{l=2L+1}^{2N}J(H_l)\, J^{(0)}(H_{2L},{S},H_{2L+1})\nn
&&+\mathcal{O}(\tau^1).\Label{Eq:Type3II}
\eea
Here, $J^{(0)}(H_{2L},{S},H_{2L+1})$  is the leading order of sub-current $J(H_{2L},{S},H_{2L+1})$ and can be further recursively written out by \eqref{Eq:Type-3b}. Apparently, there always exist one term in the expression of $J^{(0)}(H_{2L},{S},H_{2L+1})$:
         \bea
       -\left(-{1\over 2F^2}\right)^M \,J^{(0)}(H_{2L},{S_1})\prod_{j=2}^{2M}J^{(0)}({S_j})J^{(0)}({S_{2M+1}},H_{2L+1}),
         \eea
         %
{which together with \eqref{Eq:Type3II} produces \eqref{Eq:Type3I} with an opposite sign precisely. Thus such term from \figref{OddNumberSoft}. (B) cancel with \figref{OddNumberSoft}. (A) for given $\{H_I, {S_{1}}\},\{{S_2}\},\cdots,\{{S_{2M}}\},\{{S_{2M+1}},H_{I+1}\}$. After considering all such cancelations, we arrive the expected behavior \eqref{Eq:Type-3a}.}

\subsubsection*{Proof of \eqref{Eq:Type-3b}}

{Eq. (\ref{Eq:Type-3b}) gives the behavior with $a=2l$. In this case, typical diagrams are also given by \figref{OddNumberSoft}. (A) and (B) but with $a=2l$.
There are odd number of hard pions appear before/after  the first/last soft pion. As already discussed in the proof of \eqref{Eq:Type-3b},
Only the \figref{OddNumberSoft}. (A) and (B) with both $\{H_{I}\}$ and $\{H_{I+1}\}$ contain odd number hard pions provide nonzero contributions.}
Consequently, $I$ in \figref{OddNumberSoft}. (A) and (B) should be odd, i.e., $I=2L+1$ for some $L$.

For any {diagram given by \figref{OddNumberSoft}. (A)} with $L>0$ or $L<N-1$, one can always find a corresponding cancelation from  \figref{OddNumberSoft}. (B), as shown in {the proof of \eqref{Eq:Type-3a}}.
A subtle situation comes from $L=0$ and $N=1$.
In this case, all hard pions {in \figref{OddNumberSoft}. (A)} appearing before/after the first/last soft pion {belong to} a single sub-current $J(H_1,S_1)$/$J(S_{2M+1},H_2)$.
{There is no corresponding \figref{OddNumberSoft}. (B) can be found.}
 {For each \figref{OddNumberSoft}. (A) of this special case, the leading orders of propagator and vertex are
${i\over \left(P_{H_1}+P_{H_2}\right)^2}$ and $i\left(-{1\over 2F^2}\right)^{M}\left(P_{H_1}+P_{H_2}\right)^2$,
while the sub-currents contribute
\bea
-\left(-{1\over 2F^2}\right)^{M}J^{(0)}(H_1,{S_1})\prod_{i=2}^{2M}J^{(0)}({S_i})J^{(0)}({S_{2M+1}},H_2),
\eea
which is exactly one term in \eqref{Eq:Type-3b}.}
After summing all diagrams of this special case, we get the right hand side of  \eqref{Eq:Type-3b}.

\section{Amplitudes with  adjacent soft pions of even number } \label{Sec:EvenSoft}

\begin{figure}
  \centering
 \includegraphics[width=0.77\textwidth]{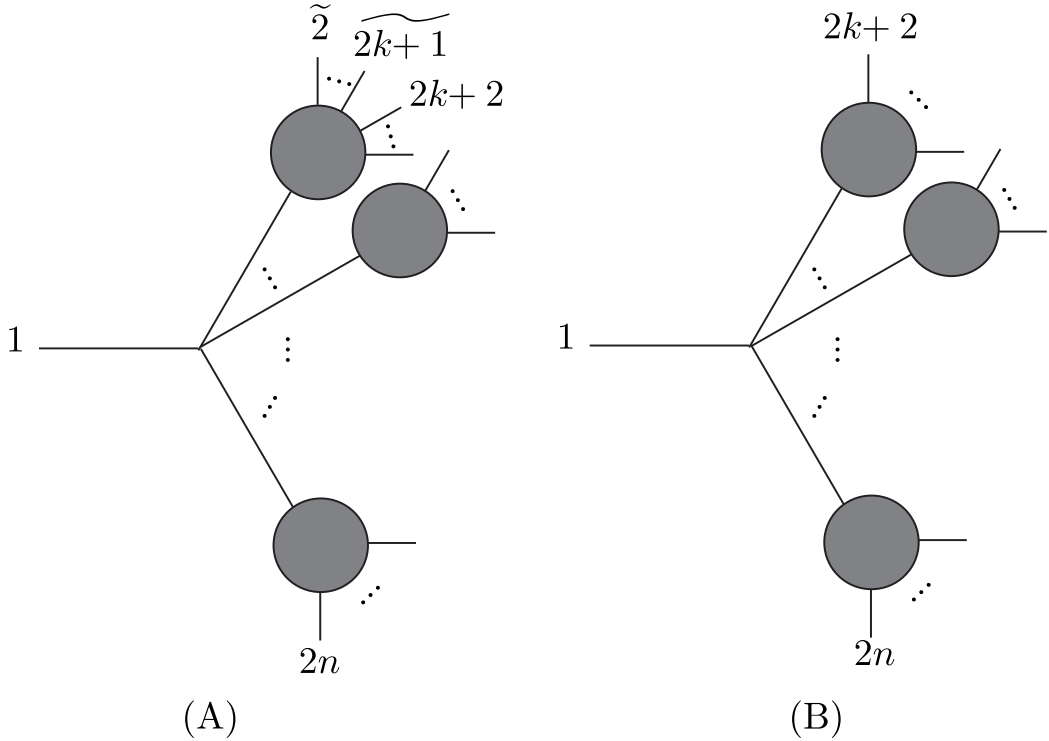}
 \caption{ (A) is a typical diagram for $J(\W 2, \cdots, \W {2i+1}, 2i+2,\cdots)$ with all soft pions living in a same sub-current.  When the soft factor $J^{(0)}(\W 2,\cdots,\W {2k+2}, 2k+2)$ is extracted, the diagram (A) becomes diagram (B).} \label{EvenSoftFig1}
\end{figure}
\begin{figure}
  \centering
 \includegraphics[width=0.9\textwidth]{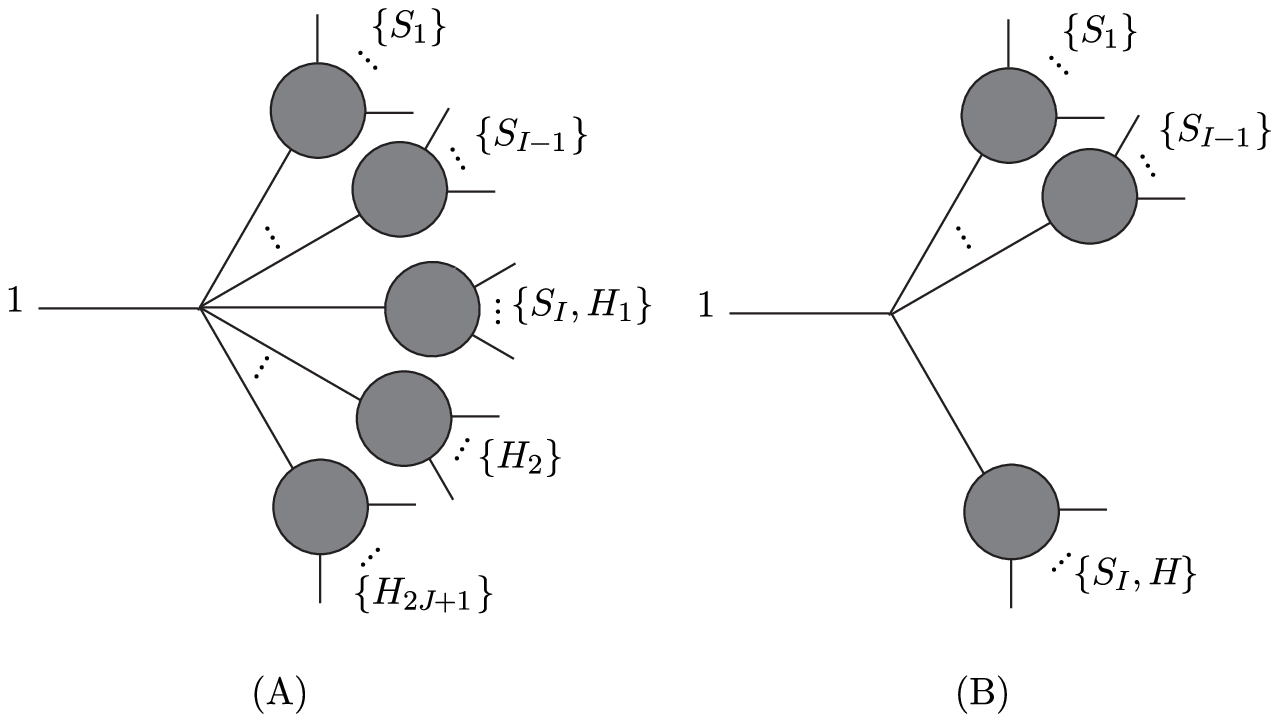}
 \caption{Tpypical diagrams for $J(\W 2, \cdots, \W {2i+1}, 2i+2,\cdots)$ with soft pions belong to more than one sub-currents: (A) is a diagram with hard pions belong to more than one sub-current; (B) is a diagram with all hard pions belong to a same sub-current. } \label{EvenSoftFig2}
\end{figure}
%

{Now let us consider the behavior of amplitude $A(1,\cdots,a,\W {a+1},\cdots,\W {a+2i},a+2i+1,\cdots,2n)$  with adjacent soft pions of even number and prove \eqref{eq:EvenAdjSoft}.} {If the amplitude contains two hard pions, one can use both on-shell recursion and off-shell Berends-Giele recursion directly to prove that the $\tau^0$ order should vanish.
In this section, we consider the nontrivial case with at least four hard pions.} First of all, we {prove} the {adjacent} even soft limits of {Berends-Giele} currents ({defined by \eqref{eq:B-G}}) instead of amplitudes inductively.
With the currents' results at hand, the amplitude version is derived eventually, as \eqref{eq:EvenAdjSoft} in the introduction. Since we will use Berends-Giele recursion relations in the following discussion, different relative positions between the soft pions and the off-shell line should be considered.


\subsection{One soft pion is adjacent to the off-shell line}

If one soft pion is adjacent to the off-shell line, the {Berends-Giele} current  with soft pions of even number  has the form $J(\W 2,\dots,\W{2i+1},2i+2,\dots,2n)$.
{When $i=n-1$, the current only contain one hard pion, one can easily check that this current is of $\tau^0$ order for $n=2$ and prove that $J(\W 2,\dots,\W{2n-1}, 2n)$ is also of order $\tau^0$ by Berends-Giele recursion for $n>2$.}
{If $i<n-1$, the current $J(\W 2,\dots,\W{2i+1},2i+2,\dots,2n)$ in the  soft limit $\tau\to 0$} should obey the following identity
\bea
J(\W 2,\cdots,\W{2i+1},2i+2,\cdots,2n)=J^{(0)}(\W 2,\cdots,\W{2i+1}, 2i+2)J(2i+2,\cdots,2n), \Label{eq:EvenSoftAdjType1}
\eea
where $J^{(0)}(\W 2,\cdots,\W{2i+1}, 2i+2)$ is the $\tau^0$ order term of $J(\W 2,\cdots,\W{2i+1}, 2i+2)$.
To {prove} this identity, we classify the diagrams in Berends-Giele recursion \eqref{eq:B-G} with a criterion that all soft pions are in a single sub-current or not.

\begin{itemize}
\item {\bf \textit{All soft pions {are} in a single sub-current}}

A typical diagram where all soft pions are located in a single sub-current is expressed in \figref{EvenSoftFig1}. (A).
Since a sub-current can only contain odd number of external pions, {there must be even number soft pions and odd number hard pions in the first sub-current.}
From inductive assumption, the {sub-current $J(\W 2, \cdots, \W {2i+1}, 2i+2,\cdots)$ satisfies \eqref{eq:EvenSoftAdjType1} when $\tau\to 0$.
When the leading soft factor $J^{(0)}(\W 2,\dots, \W{2i+1}, 2i+2, \cdots)$  is extracted out, \figref{EvenSoftFig1}. (A) becomes \figref{EvenSoftFig1}. (B).
Thus we established a one to one correspondence between diagrams (A) and (B) in \figref{EvenSoftFig1}. Notice that} the summation of all diagrams (B) is exactly the Berends-Giele current of only hard pions, i.e.,  $J(2i+2,\cdots,2n)$.
All together, the collection of this type {diagrams} with all soft pions in a single sub-current result in \eqref{eq:EvenSoftAdjType1}.

\item {\bf \textit{More than one sub-currents {contain} soft pions}}

A typical diagram of this type is {given by} \figref{EvenSoftFig2}. (A).
The ordered set $\{S_I, H_1\}$ in \figref{EvenSoftFig2}. (A) contains both soft and hard external pions.
{If there are odd number of soft pions in  $\{S_I, H_1\}$, the $\tau^0$ order of the sub-current $J(S_I,H_1)$ is zero as already stated in \secref{off-shell-odd}.}
 Thus $\{S_I, H_1\}$ can only contain soft pions of even number. {The boundary case of \figref{EvenSoftFig2}. (A), where $S_I$ is empty and $\{S_I, H_1\}=\{H_I\}$ only consists of hard pions, must also be taken into account.
The $\tau^0$ ($\tau\to 0$) order of  \figref{EvenSoftFig2}. (A) evaluates (for convenience we omit the coupling constants which do not affect our discussions)}
 \bea
 {i\over \left(P_{H_1}+P_{H_2}+\dots+P_{H_{2J+1}}\right)^2}\,i\left(\Sl_{l=0}^{J}P_{H_{2l+1}}\right)^2\left[\prod_{l=1}^{I-1}J^{(0)}(S_i)\right]J(S_I,H_1)\Big|_{\tau\to 0}\prod_{l=2}^{2J+1} J(H_l).
 \eea
From inductive assumption, we can use \eqref{eq:EvenSoftAdjType1} to express $J(S_I,H_1)\Big|_{\tau\to 0}$ as
\bea
J(S_I,H_1)\Big|_{\tau\to 0}=J^{(0)}(S_I,2k+2)J(H_1),
\eea
where $2k+2$ is the first hard {pion} in $\{H_1\}$.
Thus the leading order of \figref{EvenSoftFig2}. (A) behaves as
\bea
{i\over \left(P_{H_1}+P_{H_2}+\dots+P_{H_{2J+1}}\right)^2}\,i\left(\Sl_{l=0}^{J}P_{H_{2l+1}}\right)^2\left[\prod_{l=1}^{I-1}J^{(0)}(S_i)\right]J^{(0)}(S_I,2k+2)\prod_{l=1}^{2J+1} J(H_l). \Label{(A)InFig2}
\eea

{The special case with $J=0$, where all hard pions belong to a single sub-current as shown by \figref{EvenSoftFig2}. (B), should be studied separately.
One can find cancellations between diagrams (A) and (B) in \figref{EvenSoftFig2}.}
To see this, we compute the leading behavior of the special case \figref{EvenSoftFig2}. (B), with \eqref{eq:EvenSoftAdjType1} for lower-point sub-current {and obtain}
\bea
{i\over \left(P_{H_1}+\cdots+P_{H_{2J+1}}\right)^2}\,i\left(\Sl_{j=1}^{2J+1}P_{H_{j}}\right)^2\left[\prod_{l=1}^{I-1}J^{(0)}(S_i)\right]J^{(0)}(S_I,2k+2) J(2k+2,\cdots,2n). \Label{(B)InFig2}
\eea
According to Berends-Giele recursion \eqref{eq:B-G}, the current $J(2k+2,\cdots,2n)$ can be expanded as
\bea
J(2k+2,\cdots,2n)={i\over \left(\Sl_{j=1}^{2J+1}P_{H_j}\right)^2}\Sl_{\text{Divisions}}i\left(\Sl_{l=0}^{J }P_{H_{2l+1}}\right)^2\prod\limits_{l=1}^{2J+1}J(H_l).
\eea
Here we summed over all possible divisions of hard pions $\{2k+2,\dots,2n\}\to \{H_1\},\cdots,\{H_{2J+1}\}$, where each subset contains odd number of hard pions.
Apparently, \eqref{(A)InFig2} always cancels with {the corresponding division $\{H_1\},\cdots,\{H_{2J+1}\}$ of \eqref{(B)InFig2}}, thus this type of diagrams do not contribute at all.
\end{itemize}

Summing all two types of diagrams, the proof of the leading behavior \eqref{eq:EvenSoftAdjType1} of current with one soft pion adjacent to the off-shell line is {completed}.
This conclusion will be used as a known fact in our following discussion about the situation that {no soft pion is adjacent to the off-shell leg.}

%
\begin{figure}
  \centering
 \includegraphics[width=0.9\textwidth]{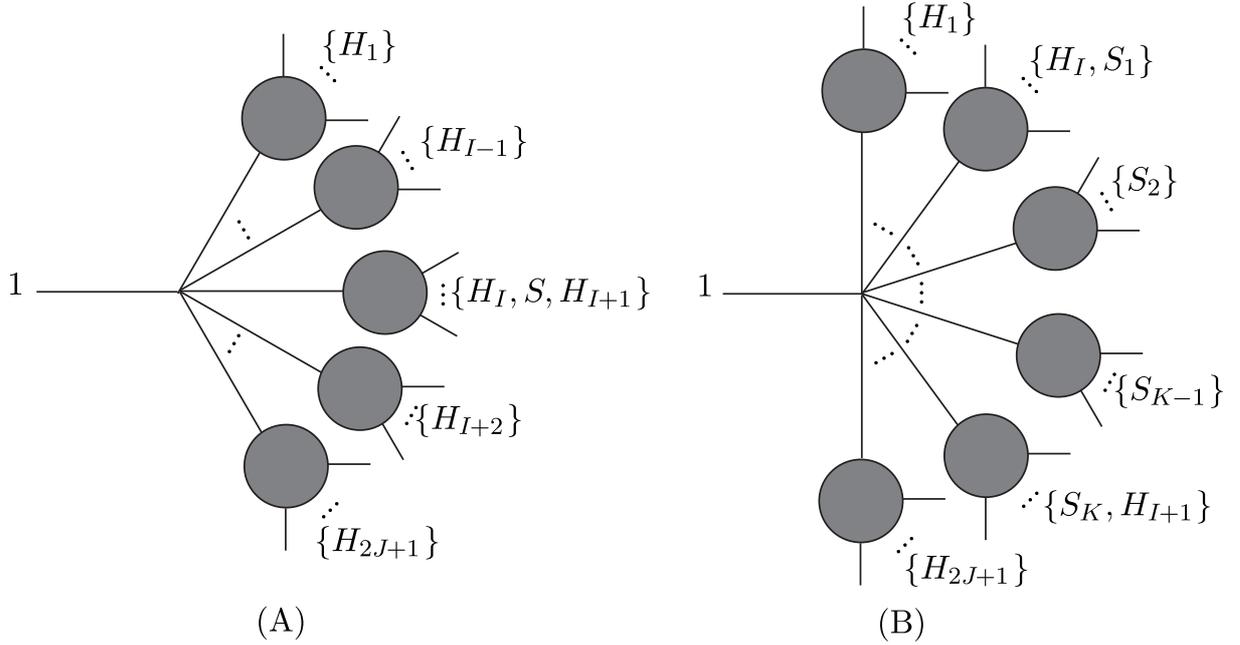}
 \caption{Typical diagrams of $J(2,\cdots,a,\W{a+1},\cdots,\W{a+2i},{a+2i+1},\cdots,2n)$: (A) is a diagram with all soft pions living in a same sub-current; (B) is a diagram with soft pions belong to more than one diagram. } \label{EvenSoftFig3}
\end{figure}

\subsection{All soft pions are nonadjacent to the off-shell line}
{Now let us study the behavior of Berends-Giele current  $J( 1,\cdots,a,\W{a+1},\cdots,\W{a+2i},{a+2i+1},\cdots,2n)$ where no soft pion is adjacent to the off-shell leg and show that}
\bea
J(2,\cdots,a,\W{a+1},\cdots,\W{a+2i},{a+2i+1},\cdots,2n)=\mathbb{S}^{(0)}_{a,a+2i+1}J(1,\cdots,a,a+2i+1,\cdots,2n)+\mathcal{O}(\tau).\Label{eq:EvenSoftAdjType2}
\eea
{Here} the soft factor $\mathbb{S}^{(0)}_{a,a+2i+1}$ is defined by \eqref{eq:EvenAdjSoftFactor}.

To prove the above behavior, {we} expand $J(2,\cdots,a,\W{a+1},\cdots,\W{a+2i},{a+2i+1},\cdots,2n)$ by Berends-Giele recursion.
According to {whether} the two hard pions $a$ and $a+2i+1$ are in a same sub-current or not,
the diagrams in Berends-Giele recursion expressions {are classified} into two types, {shown by \figref{EvenSoftFig3}. (A) and (B) respectively.}
{Substituting} \eqref{eq:EvenSoftAdjType2} into the sub-current $J(H_L,S,H_R)$ and {extracting} the soft factor $\mathbb{S}^{(0)}_{a,a+2i+1}$ out, {we find that} \figref{EvenSoftFig3}. (A) becomes a typical diagram of the Berends-Giele recursion expression of $J(1,\cdots,a,a+2i+1,\cdots,2n)$, where $a$, $a+2i+1$ are in a same sub-current.

{On the other hand, when \eqref{eq:EvenSoftAdjType1} is applied} to sub-currents $J(H_I,S_1)$ and $J(S_K,H_{I+1})$, the leading behavior of  \figref{EvenSoftFig3}. (B) can be written as
\bea
J^{(0)}(a,S_1)J^{(0)}(S_2)\cdots J^{(0)}(S_K,a+2i+1)\left[{i\over \left(P_{H_1}+\cdots+P_{H_{2J+1}}\right)^2}i\left(\Sl_{l=0}^JP_{H_{2l+1}}\right)^2\prod\limits_{b=1}^{2J+1}J(H_j)\right].
\eea
{For diangrams shown by \figref{EvenSoftFig3}. (B) with} given division of hard pions $\{H_1\},\cdots,\{H_{2J+1}\}$, once summing over all possible
$J^{(0)}(a,S_1)J^{(0)}(S_2)\cdots J^{(0)}(S_K,a+2i+1)$, one can get a diagram, in which all external lines are hard pions and the pions $a$, $a+2i+1$ are separated in two different sub-currents, multiplied by a soft factor \eqref{eq:EvenAdjSoftFactor}.

To conclude, up to a soft factor  \eqref{eq:EvenAdjSoftFactor}, the (A) type of diagrams produce diagrams in $J(1,\cdots,a,a+2i+1,\cdots,2n)$  with $a$, $a+2i+1$ in a same sub-currents while the (B) type of diagrams produce those diagrams in $J(1,\cdots,a,a+2i+1,\cdots,2n)$ with  $a$, $a+2i+1$ in two different sub-currents. Summing all possible (A) and (B) types of diagrams in \figref{EvenSoftFig3} together and {putting coupling constants back}, we arrive \eqref{eq:EvenSoftAdjType2}.

\subsection{The behavior of on-shell amplitudes with {even number of} adjacent soft pions}
\begin{figure}
  \centering
 \includegraphics[width=0.9\textwidth]{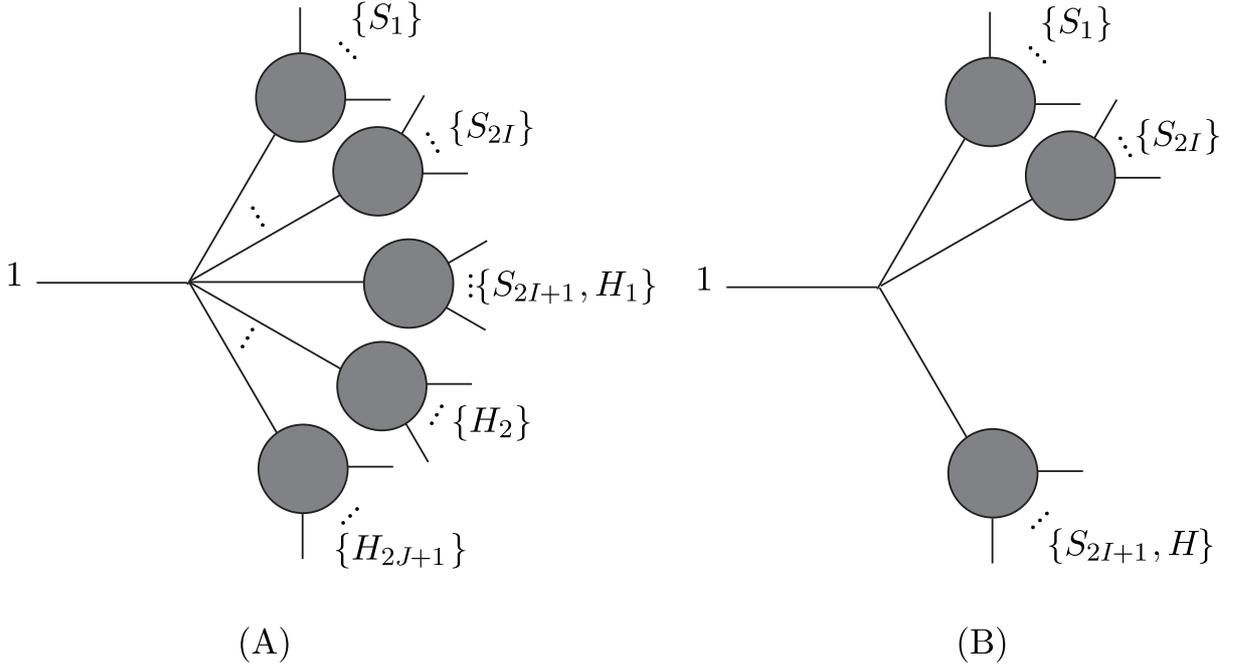}
 \caption{When the pions $1$ is set to on-shell, there are two types of diagrams for the amplitude $A(1,\W 2,\cdots,\W{2i+1},2i+2,\cdots, 2n)$: (A) stands for diagrams with hard pions belong to different sub-currents; (B) stands for diagrams with all hard pions living in a same sub-current.} \label{EvenSoftFigure4}
\end{figure}

Now we are ready to take the on-shell limit.
When $-P^2_{2,2n}$ is multiplied to the leading term of the currents $J(2,\cdots,a,\W {a+1},\cdots,\W{a+2i},a+2i+1,\cdots,2n)$  in \eqref{eq:EvenSoftAdjType2} with no soft pion adjacent to the off-shell line, we get $\mathbb{S}^{(0)}_{a+1,a+2i}A(1,\cdots,a,a+2i+1,\cdots,2n)$, which is the expected {leading order} multi-soft behavior \eqref{eq:EvenAdjSoft}.

The on-shell limit of currents with one soft pion adjacent to the off-shell line is more subtle. {Typical diagrams of this case are shown by \figref{EvenSoftFigure4}. (A) and (B).}
Apparently, if $-P^2_{2,2n}$ is multiplied to \eqref{eq:EvenSoftAdjType1} directly, the expected behavior \eqref{eq:EvenAdjSoft} of on-shell amplitudes can not be obtained.
{This is because when we consider the on-shell limit $k_1\to 0$ of  \figref{EvenSoftFigure4}. (B), the propagator of the  off-shell leg in the sub-current  $J(S_{2I+1}, H)$, which contain all hard pions in it, should be divergent in the soft limit $\tau\to 0$.}
In the \appref{sec:appendix-boundary}, the semi-amplitude $A^*(1^*,2,\dots,2n)\equiv -P_{2,2n}^2J(2,\dots,2n)$ instead of Berends-Giele current is used to study the multi-soft behavior in this case.
This semi-amplitude is nothing but \textit{the sum of all possible Feyman diagrams where $2, \cdots, 2n$ are on-shell}.
We will prove that a semi-amplitude $A^*\bigl(k_1+\tau \Sl_{j=2}^{2m-1}q_j,\W {2m},\W {2m+1},\cdots,\W {2i+1}, 2i+2 \cdots,2n\bigr)$ obeys the following identity
\bea
&&A^*\bigl(k_1+\tau \Sl_{j=2}^{2m-1}q_j,\W {2m},\W {2m+1},\cdots,\W {2i+1}, 2i+2 \cdots,2n\bigr)\nn
&=&\mathbb{S}^{*(0)}_{k_1+\tau \Sl_{j=2}^{2m-1}q_j,2i+2}A\left(1,2i+2,\cdots,2n\right)+\mathcal{O}(\tau),
\Label{eq:EvenSoftAdjType3}
\eea
{Here we use a star * to {\blue emphasize} that the semi-amplitude and soft factors, which are expressed by products of $\tau^0$ order of sub-currents, are extended to off-shell case by just allowing external off-shell legs in the Feynman rules \eqref{eq:Feyn-rules}.}
When $m=1$, the momentum of the first leg becomes the on-shell momentum $k_1$, we get the expected soft behavior of amplitude $A(1,\W 2,\cdots,\W {2i+1},2i+2,\cdots,2n)=\mathbb{S}^{(0)}_{1,2i+2}A(1,2i+2,\cdots,2n)+\mathcal{O}(\tau)$.
Details can be found in \appref{sec:appendix-boundary}.

\subsection{Quadruple soft limits}

{\red When we have two adjacent soft pions, the behavior \eqref{eq:EvenAdjSoft}, having three possible divisions in the soft factor \eqref{eq:EvenAdjSoft}, precisely reproduces the double soft behavior \cite{ArkaniHamed:2008gz,Kampf:2013vha, Cachazo:2015ksa, Du:2015esa,Low:2015ogb}.} let us have a look at the explicit expression of quadruple soft limit with adjacent soft pions. By substituting the explicit expression of $J^{(0)}$'s corresponding to the divisions \eqref{eq:4SoftDivisions} into the soft factor \eqref{eq:EvenAdjSoftFactor} with four adjacent soft pions, we obtain
{\bea
&&A_8(1,\W 2,\W 3,\W 4,\W 5, 6,\cdots,2n)\nonumber\\
 &=&{1\over 4 F^4}\,A_4(1,6,\cdots,2n) \,\,\bigg[
-{s_{2,4}\over s_{234}}{q_5\cdot k_6\over (q_2+q_3+q_4+q_5)\cdot k_6}+{s_{2,4}\over s_{234}}{q_4\cdot k_1\over (q_2+q_3+q_4+q_5)\cdot k_1}
 \nonumber\\
 &&\quad+{s_{3,5}\over s_{345}}{q_2\cdot k_6\over (q_2+q_3+q_4+q_5)\cdot k_6}-{s_{3,5}\over s_{345}}{q_2\cdot k_1\over (q_2+q_3+q_4+q_5)\cdot k_1}\nonumber\\
 &&\quad
 -{q_2\cdot k_6\over (q_2+q_3+q_4+q_5)\cdot k_6}\,{q_5\cdot k_6\over (q_4+q_5)\cdot k_6}
  -\,{q_2\cdot k_1\over (q_2+q_3)\cdot k_1}\,{q_5\cdot k_1\over (q_1+q_2+q_3+q_4)\cdot k_1}\nonumber\\
  &&\quad+{q_5\cdot k_6\over (q_4+q_5)\cdot k_6}\,{q_2\cdot k_1\over (q_2+q_3)\cdot k_1}\bigg].
\eea}
The sextuple soft factor is also derived but with a long expression, we put it in  \appref{sextuple}.

\section{Amplitudes with nonadjacent soft blocks}\label{sec:NonadjacetBlocks}
In this section, we generalize the discussions in \secref{Sec:OddSoft} and \secref{Sec:EvenSoft} to amplitudes containing nonadjacent soft blocks as
\bea &&A\bigl(1,\cdots,a_1,\{\W{a_1+1},\cdots, \W{a_1+i_1}\}, a_1+i_1+1, \cdots,a_2,\{\W{a_2+1},\cdots, \W{a_2+i_2}\}, a_2+i_2+1,\nn
&&~~~~~~\cdots, a_r,\{\W{a_r+1},\cdots, \W{a_r+i_r}\}, a_r+i_r+1,\cdots \bigr).\nonumber\eea
%
%
\begin{itemize}
\item If at least one soft block, e.g., $\{\W{a_j+1},\cdots, \W{a_j+i_j}\}$ ($1\leq j\leq r$ ),  contains soft pions of odd number, the $\tau^0$ order of amplitude must vanish.
This can be directly seen when replacing some of the hard pions in $\{2,\cdots,a_j-1\}$ and $\{a_j+i_j+2,\cdots, 2n\}$ of $J\bigl(2\cdots, a_j, \W {a_j+1},\cdots,\W {a_j+i_j}, a+i_j+1,\cdots,2n\bigr)$ by some soft ones and repeating the proof procedure in \secref{Sec:OddSoft}.

\item If every soft block contains soft pions of even number, the proof given in \secref{Sec:EvenSoft} can be applied.
In particular, through a similar proof in \secref{Sec:EvenSoft},
the following iterative expression for amplitudes with soft blocks of even order is obtained
\bea
&&A\bigl(1,\cdots,a_1,\{\W{a_1+1},\cdots, \W{a_1+2i_1}\}, a_1+2i_1+1, \cdots,a_2,\{\W{a_2+1},\cdots, \W{a_2+2i_2}\}, a_2+2i_2+1,\nn
&&~~~~~~\cdots, a_r,\{\W{a_r+1},\cdots, \W{a_r+2i_r}\}, a_r+2i_r+1,\cdots \bigr)\nn
&=&\mathbb{S}^{(0)}_{a_1,a_1+2i_1+1} A^{(0)}\bigl(\cdots,a_2,\{\W{a_2+1},\cdots, \W{a_2+2i_2}\}, a_2+2i_2+1,\nn
&&~~~~~~~~~~~~~~~~~~~~~~\cdots, a_r,\{\W{a_r+1},\cdots, \W{a_r+2i_r}\}, a_r+2i_r+1,\cdots\bigr)+\mathcal{O}(\tau).
\eea
{\red The special case with only two hard pions and two even-number soft blocks, which has vanishing $\tau^0$ order behavior, is not included in the above discussion.}
\end{itemize}
Repeating this iterative procedure, we finally get the leading behavior
\bea
&&A\bigl(1,\cdots,a_1,\{\W{a_1+1},\cdots, \W{a_1+2i_1}\}, a_1+2i_1+1, \cdots,a_2,\{\W{a_2+1},\cdots, \W{a_2+2i_2}\}, a_2+2i_2+1,\nn
&&~~~~~~\cdots, a_r,\{\W{a_r+1},\cdots, \W{a_r+2i_r}\}, a_r+2i_r+1,\cdots \bigr)\nn
&=&\prod\limits_{j=1}^r\mathbb{S}^{(0)}_{a_j,a_j+2i_j+1} A\bigl(\cdots,a_1, a_1+2i_1+1, \cdots,a_2,a_{2+2i_2+1},\cdots, a_r, a_r+2i_r+1,\cdots\bigr)+\mathcal{O}(\tau).
\eea
This result is always the same no matter consecutively or simultaneously extracting soft factors.

\section{Conclusion}\label{Sec:Conclusion}

 In this paper, all leading order {multi-soft} behaviors of tree amplitudes in NLSM are discussed.
 We showed that the leading order behavior of an amplitude with odd number all-adjacent soft pions vanishes, while the leading behavior of an amplitude  with all-adjacent even number of soft pions is described by sum of products of {$\tau^0$ order} Berends-Giele sub-currents.
 Furthermore, we generalized our discussions to amplitudes with nonadjacent blocks of soft pions:
The amplitudes containing at least one block with soft pions of odd number {have to} vanish at the leading order;
If each block contains soft elements of even number, the leading order is given by products of soft factors for those blocks.

Several questions are worth further studying:
a) How to explain the multi-soft factor, especially the relations with the commutations of the generators for the broken and unbroken symmetry;
b) It would be interesting to have a look at the sub-leading order soft behavior, which might contain more insights about the global symmetry breaking;
c) We only discussed global symmetry breaking in this paper, then multi-soft behavior of supersymmetry as well as conformal symmetry breaking is another interesting topic.
{\blue d) It is worthy studying these soft behaviors of NLSM from other perspectives, e.g, Cachazo-He-Yuan formula \cite{Cachazo:2013gna,Cachazo:2013hca,Cachazo:2013iea,Cachazo:2014xea,Cachazo:2016njl} and abelian Z-theory\cite{Carrasco:2016ldy}.}

\acknowledgments
We would like to thank Rutger Boels, Clifford Cheung, Ian Low and Congkao Wen for helpful suggestions. YD would like to acknowledge National Natural Science Foundation of
China under Grant Nos. 11105118, 111547310, as well as the 351 program of Wuhan University. HL is supported by the German
Science Foundation (DFG) within the Collaborative Research Center 676 ``Particles, Strings and the Early
Universe".

\appendix
\begin{figure}
\centering
\includegraphics[width=1\textwidth]{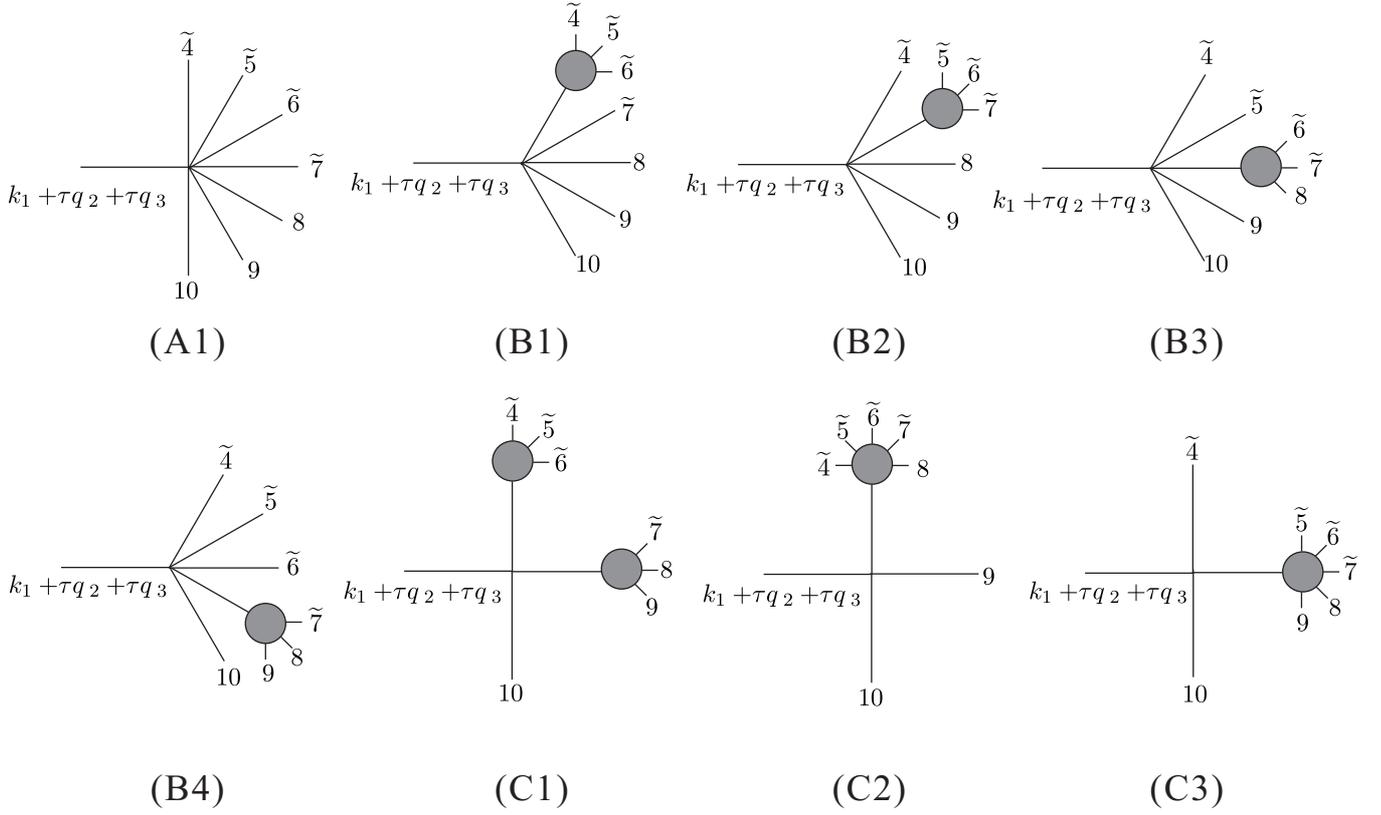}
\caption{All diagrams for semi-amplitude $A^*(k_1+\tau q_2+\tau q_3,\W4,\W5,\W6,\W 7,8,9,10)$ with hard pions belong to more than one sub-current.}  \label{Appendix1}
\end{figure}

\appendix

\begin{figure}
\centering
\includegraphics[width=1\textwidth]{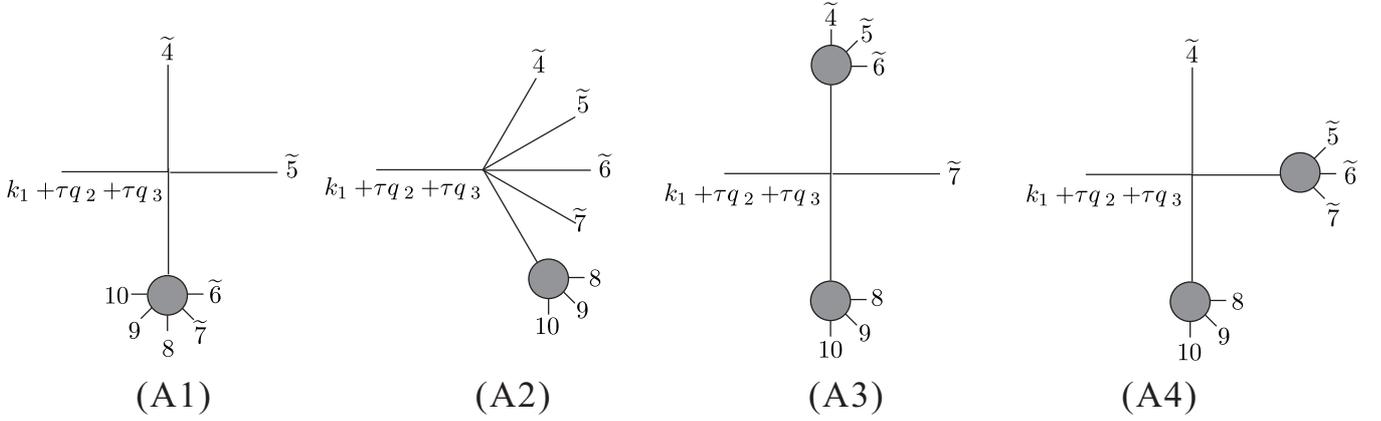}
\caption{All diagrams for semi-amplitude $A^*(k_1+\tau q_2+\tau q_3,\W4,\W5,\W6,\W 7,8,9,10)$ with all hard pions belong to a same sub-current.}  \label{Appendix2}
\end{figure}
\section{The proof of \eqref{eq:EvenSoftAdjType3}}\label{sec:appendix-boundary}

\subsection{An example}
Consider an amplitude $A(1,\W2,\W3,\W4,\W5,\W6,\W7,8,9,10)$ and choose the momentum of the pion $1$ to be expressed through momentum conservation, in other words, leg $1$ is chosen to be off-shell in the Berends-Giele recursions.
While Berends-Giele recursions are applied to rewrite this amplitude, a type of sub-currents where all hard lines are included in a same sub-current emerges, e.g., the sub-current $J(\W4,\W5,\W6,\W 7,8,9,10)$.
Since the propagator ${i\over P^2_{4,10}}={i\over (k_1+\tau q_2+\tau q_3)^2}$ behaves as ${1\over \tau}$ when taking $\tau\to 0$, the sub-current seems to be divergent.
Nevertheless, this divergence will disappear when we turn to the full amplitude.
Thus, it's more convenient to consider a semi-amplitude $A^*(k_1+\tau q_2+\tau q_3,\W4,\W5,\W6,\W 7,8,9,10)=\left(-P_{(k_1+\tau q_2+\tau q_3)}^2\right)J(\W4,\W5,\W6,\W 7,8,9,10)$ instead.
This semi-amplitude consist of all possible Feynman diagrams, but with some of external leg are off-shell.
{For example the} semi-amplitude $A^*(k_1+\tau q_2+\tau q_3,\W4,\W5,\W6,\W 7,8,9,10)$, the first leg $k_1+\tau q_2+\tau q_3$ is not on-shell unless taking a soft limit.
This semi-amplitude can be expanded {through} the Berends-Giele recursions and classified into {the following} two types of diagrams\footnote{Again, we omit coupling constants and put them back in the final expression.}.
\begin{itemize}

\item Diagrams where hard pions are included in more than one sub-currents {(see \figref{Appendix1})}:

{Since diagrams \figref{Appendix1}. (B4), (C1) and (C3) contain sub-currents with odd number of adjacent soft pions, they must have vanishing $\tau^0$ order in the soft limits $\tau\to 0$.}
The leading soft behavior of other five diagrams apparently produces
\bea
A(1,8,9,10)&&\Bigl[J^{(0)}(\W 4,\W 5,\W 6,\W 7,8)+J(\W 4,\W 5,\W 6)J^{(0)}(\W 7)J^{(0)}(8)+J^{(0)}(\W 4)J^{(0)}(\W 5,\W6,7) J^{(0)}(8)\nn
&&~~~~~~~~~~~~+J^{(0)}(\W 4) J^{(0)}(\W 5)J^{(0)}(\W 6,\W 7, 8)+J^{(0)}(\W 4)J^{(0)}(\W 5)J^{(0)}(\W 6)J^{(0)}(\W 7)J^{(0)}(8)\Bigr].\Label{Eq:10PTSoft1}
\eea

\item Diagrams where all hard pions are included in a single sub-current:

This type is more subtle, consisting of diagrams given by \figref{Appendix2}.
In \figref{Appendix2}. (A1), there is a semi-amplitude $A^*(k_1+\tau q_2+\tau q_3+\tau q_4+\tau q_5, \W 6, \W 7, 8, 9, 10)$.
As iterative assumption, the behavior of lower-point is known, i.e., the leading behavior of this semi-amplitude satisfies \eqref{eq:EvenSoftAdjType3}, as
\bea
&&A^{*(0)}\Bigl(k_1+\tau \Sl_{i=2}^5 q_i, \W 6, \W 7, 8, 9, 10\Bigr)\nn
&=&\Biggl[A^*\Bigl(k_1+\tau \Sl_{i=2}^5 q_i,\W 6, \W 7, -(k_1+\tau \Sl_{i=2}^7 q_i)\Bigr){-i\over(k_1+\tau \Sl_{i=2}^7 q_i)^2 }\Biggr]\Bigg|_{\tau\to 0}J^{(0)}(8)A(1,8,9,10)\nn
&&+\Biggl[J^{(0)}(\W 6,\W 7,8)+J^{(0)}(\W 6)J^{(0)}(\W 7)J^{(0)}(8)\Biggr]A(1,8,9,10). \Label{Eq:AppEg1}
\eea
where {the subcurrents $J(\W 6)$, $J(\W 7)$, $J(8)$ and the semi sub-current $J(k_1+\tau \Sl_{i=2}^5q_i)$, with only one element, is $1$. Thus all their $\tau^0$ order should be $1$.}
Sum the first term in \eqref{Eq:AppEg1} together with contributions from diagrams (A2), (A3), (A4), {we arrive}
\bea
&&A(1,8,9,10)\nn
&&\times\Biggl[iV_4(k_1+\tau q_2+\tau q_3, \W 4, \W 5, -(k_1+\tau \Sl_{i=2}^5 q_i)){i\over \Bigl(k_1+\tau \Sl_{i=2}^5 q_i\Bigr)^2}A^*\Bigl(k_1+\tau \Sl_{i=2}^5 q_i,\W 6, \W 7, -(k_1+\tau \Sl_{i=2}^7 q_i)\Bigr)\nn
 &&~~~~~~~~~~+iV_4(k_1+\tau q_2+\tau q_3, \tau\Sl_{i=4}^6 q_i, \W 7, -(k_1+\tau \Sl_{i=2}^7q_i))J(\W 4,\W 5,\W 6)\nn
 &&~~~~~~~~~~~~~~~~+iV_4(k_1+\tau q_2+\tau q_3, \W 4, \tau\Sl_{i=5}^7q_i, -(k_1+\tau \Sl_{i=2}^7q_i))J(\W 5,\W 6,\W 7)\nn
 &&~~~~~~~~~~~~~~~~~~+iV_6(k_1+\tau q_2+\tau q_3, \W 4,\W 5, \W 6, \W 7,-(k_1+\tau \Sl_{i=2}^7 q_i))
 \Biggr]{-i\over\Bigl(k_1+\tau \Sl_{i=2}^7 q_i\Bigr)^2}\Biggl|_{\tau\to 0}\Label{Eq:AppEg2}
\eea
The sum of the terms in the square brackets is a semi-amplitude $A^*(k_1+\tau q_2+\tau q_3, \W 4, \W 5, \W 6, \W 7, -(k_1+\tau \Sl_{i=2}^7q_i))$, thus the above expression is $A(1,8,9,10 )J^{*(0)}(k_1+\tau q_2+\tau q_3,\W4, \W 5, \W 6, \W 7)$.
The last line in \eqref{Eq:AppEg1} contributes a
\bea
&&A(1,8,9,10)iV_4\Bigl(k_1+\tau q_2+\tau q_3, \W 4, \W 5,-(k_1+\tau\Sl_{i=2}^5q_i)\Bigr){-i\over\Bigl(k_1+\tau \Sl_{i=2}^5 q_i\Bigr)^2}\Bigg|_{\tau\to 0} \nn
&&\times\left[J^{(0)}(\W 6,\W 7,8)+J^{(0)}(\W 6)J^{(0)}(\W 7)J(8)\right]\nn
&=&A(1,8,9,10)J^{*(0)}(k_1+\tau q_2+\tau q_3, \W 4, \W 5)\left[J^{(0)}(\W 6,\W 7,8)+J^{(0)}(\W 6)J^{(0)}(\W 7)J(8)\right], \Label{Eq:AppEg3}
\eea
to the leading behavior of \figref{Appendix2}. (A1).
Thus all the diagrams in \figref{Appendix2} are summed as
\bea
&&A(1,8,9, 10)\Biggl[J^{*(0)}(k_1+\tau q_2+\tau q_3, \W 4, \W 5, \W 6, \W 7)J(8)+J^{*(0)}(k_1+\tau q_2+\tau q_3, \W 4, \W 5 )J^{(0)}(\W 6,\W 7,8)\nn
&&~~~~~~~~~+J^{*(0)}(k_1+\tau q_2+\tau q_3, \W 4, \W 5)J^{(0)}(\W 6)J^{(0)}(\W 7)J(8)\Biggr].\Label{Eq:10PTSoft2}
\eea
\end{itemize}

{Putting coupling constants back}, \eqref{Eq:10PTSoft1} and \eqref{Eq:10PTSoft2}, corresponding to \figref{Appendix1} and \figref{Appendix2} respectively, result in the expected soft behavior of the semi-amplitude $A^*(k_1+\tau q_2+\tau q_3,\W4,\W5,\W6,\W 7,8,9,10)$ shown in \eqref{eq:EvenSoftAdjType3}.

\subsection{General proof}
%
%
From the example, we find that diagrams of the Berends-Giele recursion expansion of a semi-amplitude are classified into two categories.
This pattern can be easily generalized to an arbitrary point semi-amplitude $A^*\bigl((k_1+\Sl_{i=2}^{2m-1} q_i),\W {2m},\cdots,\W {2k+1}, 2k+2 \cdots,2n\bigr)$: \textit{i)} diagrams with hard pions living in more than one sub-current,\textit{ii) }diagrams with all hard pions in a single sub-current.
Now let us discuss these two cases separately \footnote{Again, we omit coupling constants and put them back in the final expression}.
\begin{itemize}
\item[]
\textit{i)} Diagrams where on-shell hard pions $2k+2,\dots ,2n$ belong to different sub-currents, \figref{EvenSoftFigure4}. (A):

The leading soft behavior of such a diagram is
\bea
i\left(\tau\Sl_{l=0}^IQ_{S_{2l+1}}+\Sl_{m=0}^JP_{H_{2m+1}}\right)^2\Bigg|_{\tau=0}\left[\prod_{i=1}^{2I}J^{(0)}(S_i)\right]J^{(0)}(S_{2I+1},H_1)\left[\prod_{j=2}^{2J+1}J(H_j)\right],
\eea
where $\tau Q_{S_{2l+1}}$ means summing over momenta of all soft pions in the set $S_{2l+1}$, and the current $J(H_i)$ is of order $\tau^0$ itself.
Since the sub-current $J(S_{2I+1},H_1)$ does not contain all hard pions, it satisfies the soft behavior \eqref{eq:EvenSoftAdjType1}.
The $\tau^0$ behavior of \figref{EvenSoftFigure4}. (A) is
\bea
&&J^{(0)}(S_{2I+1},2k+2)\prod_{i=1}^{2I}J^{(0)}(S_i)\,\left[i\,\left(\Sl_{m=1}^JP_{H_{2m+1}}\right)^2 \prod_{j=1}^{2J+1}J(H_j)\right].\Label{Eq:AppendixGen1}
\eea
Evidently, leading soft behavior of this type is written as
\bea
&&\Sl_{\text{Divisions}^{(A)}(S)}J^{(0)}(S_{2I+1},2k+2)\prod_{i=1}^{2I}J^{(0)}(S_i)\,\left[i\,\Sl_{J=1}^{n-1-k}\Sl_{\text{Divisions}(\{H\})}\left(\Sl_{m=1}^JP_{H_{2m+1}}\right)^2 \prod_{j=1}^{2J+1}J(H_j)\right].\nn
\eea
Here ``${\text{Divisions}^{(A)}(S)}$" stands for all possible divisions of $\W {2m}$, ..., $\W{2k+1}$.
Among those divisions, the last subset contains soft pions of even number while all other subsets contain soft pions of odd number.
The summation in square brackets is Berends-Giele recursion expansions {(i.e., the sum of all Feynman diagrams)}  of the amplitude $A(1,2k+2,\cdots,2n)$, thus the above equation finally provides
\bea
\mathbb{S}^{*(0)A}\Bigl(k_1+\tau\Sl_{i=2}^{2m-1} q_i,2k+2\Bigr)\,A\left(1,2k+2,\cdots,2n\right),
\eea
in which
\bea
\mathbb{S}^{*(0)A}\Bigl(k_1+\tau\Sl_{i=2}^{2m-1} q_i,2k+2\Bigr)\equiv \Sl_{\mathcal{D}^A}\prod_{i=1}^{\mathcal{N}(\mathcal{D}^A)}J^{(0)}(\{\mathcal{S}_{\mathcal{D}^A}^{i}\}),
\eea
where we have summed over all divisions $\mathcal{D}^A$ of $\{\W {2m}, \W {2m+1},\cdots, \W {2k+1}, {2k+2}\}$ such that each subset contains elements of odd number.

\item[] \textit{ii)} Diagrams where on-shell hard pions $2k+2,\dots, 2n$ belong to a single sub-current, \figref{EvenSoftFigure4}. (B):

Such diagram contributes a term
\bea
\left[i(\tau\Sl_{l=0}^IQ_{S_{2l+1}}+P_{H})^2\prod_{i=1}^{2I}J(S_j){i\over (k_1+\tau\Sl_{i=1}^{2m-1}q_i)^2 }A^*\Bigl((k_1+\tau\Sl_{i=1}^{2m-1}q_i),S_{2I+1},H\Bigr)\right]\Bigg|_{\tau=0}.
\eea
Considering the soft behavior of $A^*\Bigl((k_1+\tau\Sl_{i=1}^{2m-1}q_i),S_{2I+1},H\Bigr)$, from inductive assumption for lower point semi-amplitudes in \eqref{eq:EvenSoftAdjType3}, we have
\bea
A^{*(0)}\Bigl((k_1+\tau\Sl_{i=1}^{2m-1}q_i),S_{2I+1},H\Bigr)=\Sl_{\mathcal{D}'}\left(J^{*(0)}(S_{\mathcal{D}'}^1)\,\prod_{l=2}^{\mathcal{N}(\mathcal{D}')}J^{(0)}(S_{\mathcal{D}'}^l)\right)A\left(1, 2k+2,\cdots,2n\right),
\eea
where $\mathcal{D}'$ are divisions of the ordered set $\Bigl\{(k_1+\tau\Sl_{i=1}^{2m-1}q_i),S_{2I+1},2k+2\Bigr\}$.
{The sum of the leading behaviors of diagrams of \figref{EvenSoftFigure4}. (B) type contributes}
\bea
&&\Biggl\{A(1,2k+2,\cdots,2n)\Sl_{I}\Sl_{\cal D} i\Bigl(\tau \Sl_{l=0}^{I}P_{S_{2l+1}}+P_{H}\Bigr)^2\left[\prod_{j=1}^{2I}J(S_j)\right]\nn
&&\times\Biggl[\Sl_{\mathcal{D}'}{i\over {\bigl(k_1+\tau\Sl_{i=1}^{2m-1}q_i\bigr)^2 }}A^*\Bigl(\bigl(k_1+\tau\Sl_{i=1}^{2m-1}q_i\bigr),S^1_{\mathcal{D}'},-\bigl(k_1+\tau\Sl_{i=1}^{2m-1}q_i+P_{S^1_{\mathcal{D}'}}\bigr)\Bigr){i\over {\bigl(k_1+\tau\Sl_{i=1}^{2m-1}q_i+P_{S^1_{\mathcal{D}'}}\bigr)^2 }}\nn
&&~~~~\times\prod_{l=2}^{\mathcal{N}(\mathcal{D}')}J( S^l_{\mathcal{D}'})\Biggr]\Biggr\}\Biggl|_{\tau\to 0}, \Label{Eq:AppendixGen2}
\eea
 where $\mathcal{D}$ denote all possible divisions of soft pions $\{\W{2m},\W{2m+1},\cdots,\W{2k+1}\}\to\{S_1\}\cdots\{S_{2I}\},\{S_{2I+1}\}$.
 The sum over divisions $\mathcal{D}$ and $\mathcal{D}'$ can be rearranged by collecting those terms
 which have a same last soft pion in the first subset $\{S^1_{\mathcal{D}'}\}$ of  $\mathcal{D}'$ (in the example of the previous subsection, \eqref{Eq:AppEg2} is a collection of terms with the pion $7$ as the last element of $\{S^1_{\mathcal{D}'}\}$, while \eqref{Eq:AppEg3} is a collection of terms with $5$ as the last element of $\{S^1_{\mathcal{D}'}\}$ ).
 After this rearrangement,  \eqref{Eq:AppendixGen2} turns out
 \bea
 &&\Biggl\{A(1,2k+2,\cdots,2n)\Sl_{a=m}^k\Biggl[\Sl_{I=4}^{2a-2m+4}\Sl_{\mathcal{D}_{2a+1}}i\Bigl(\tau \Sl_{l=0}^{I-1}P_{S_{2l+1}}+P_{H}\Bigr)^2\Bigl[\prod_{j=1}^{I-2}J(S^j_{\mathcal{D}_{2a+1}})\Bigr]\nn
 &&\times{i\over {\bigl(k_1+\tau\Sl_{i=1}^{2m-1}q_i\bigr)^2 }}A^*\Bigl(\bigl(k_1+\tau\Sl_{i=1}^{2m-1}q_i\bigr),S^{I-1}_{\mathcal{D}_{2a+1}},-\bigl(k_1+\tau\Sl_{i=1}^{2a+1}q_i\bigr)\Bigr)\Biggr]{i\over {\bigl(k_1+\tau\Sl_{i=1}^{2a+1}q_i\bigr)^2 }}\nn
&&\times\Sl_{\mathcal{D}'\in\text{Divisions}\{2a+2,\cdots,2m+1\}}\prod_{l=1}^{\mathcal{N}(\mathcal{D}')}J( S^l_{\mathcal{D}'})\Biggr\}\Bigg|_{\tau\to 0}.
 \eea
Note that the second line in the brackets is the sub-current $J^*\bigl(S^{I-1}_{\mathcal{D}_{2a+1}},-\bigl(k_1+\tau\Sl_{i=1}^{2a+1}q_i\bigr)\bigr)$ where $-\bigl(k_1+\tau\Sl_{i=1}^{2a+1}q_i\bigr)$ is also off-shell.
All the factors given in the square brackets thus produce a semi-amplitude $A^*\bigl(k_1+\tau\Sl_{i=2}^{2m-1} q_i, \W {2m},\cdots,\W {2a+1},-\bigl(k_1+\tau\Sl_{i=1}^{2a+1}q_i\bigr) \bigr)$ which becomes a semi-current $J^*\bigl(k_1+\tau\Sl_{i=2}^{2m-1} q_i, \W {2m},\cdots,\W {2a+1}\bigr)$ if {multiplied by} a factor ${i\over {\bigl(k_1+\tau\Sl_{i=1}^{2a+1}q_i\bigr)^2 }}$.
The expression \eqref{Eq:AppendixGen2} then becomes
\bea
&&A(1,2k+2,\cdots,2n)\nn
&&\times\left[\Sl_{a=m}^kJ^{*(0)}\Bigl(k_1+\tau\Sl_{i=2}^{2m-1} q_i, \W {2m},\cdots,\W {2a+1}\Bigr)\Sl_{\mathcal{D}'\in\text{Divisions}\{2a+2,\cdots,2m+1\}}\prod_{l=1}^{\mathcal{N}(\mathcal{D}')}J( S^l_{\mathcal{D}'})\right].
\eea
Terms in the square brackets can be expressed as
\bea
\mathbb{S}^{*(0)B}\Bigl(k_1+\tau\Sl_{i=2}^{2m-1} q_i,2k+2\Bigr)\equiv \Sl_{\mathcal{D}^{B}}\prod_{l=1}^{\mathcal{N}(\mathcal{D}^B)}J^{(0)}(\{\mathcal{S}_{\mathcal{D}^B}^{l}\}).
\eea
where $\mathcal{D}^B$ stands for all possible divisions of $\Bigl\{k_1+\tau\Sl_{i=2}^{2m-1} q_i, \W {2m},\cdots,\W {2k+1}, 2k+2\Bigr\}$ such that each subset contains element of odd number and the first subset contains more than one element. Therefore \eqref{Eq:AppendixGen2} can be written as
\bea
\mathbb{S}^{*(0)B}\Bigl(k_1+\tau\Sl_{i=2}^{2m-1} q_i,2k+2\Bigr)A\left(1,2k+2,\cdots,2n\right).
\eea
\end{itemize}
Finally, {summing both types of diagrams together and plugging coupling constants back}, we  obtain the leading behavior
\bea
&&A^*\Bigl((k_1+\Sl_{i=2}^{2m-1} q_i),\W {2m},\cdots,\W {2k+1}, 2k+2 \cdots,2n\Bigr)\nn
&=&\left[\mathbb{S}^{*(0)A}\Bigl(k_1+\tau\Sl_{i=2}^{2m-1} q_i,2k+2\Bigr)+\mathbb{S}^{*(0)B}\Bigl(k_1+\tau\Sl_{i=2}^{2m-1} q_i,2k+2\Bigr)\right]A\left(1,2k+2,\cdots,2n\right)\nn
&=&\mathbb{S}^{*(0)}\Bigl(k_1+\tau\Sl_{i=2}^{2m-1} q_i,2k+2\Bigr)A\left(1,2k+2,\cdots,2n\right),
\eea
which is the expected behavior \eqref{eq:EvenSoftAdjType2}.

\section{Sextuple behavior}\label{sextuple}
The sextuple soft factor for the color-like ordered NLSM amplitudes $A(1,2, \W 3, \W4, \dots, \W 8, 9, \dots, 2n)$ is
\bea
\mathbb{S}^{(0)}_{2,8}&=&+{1\over 8F^6}{1\over 4}\left[1-\frac{s_{24}}{s_{2|3,4}}
		        -\frac{s_{97}}{s_{9|7,8}}
		        +\frac{s_{28}}{s_{2|3,8}}
		        +\frac{s_{93}}{s_{9|3,8}}\right]
\nonumber\\
&&+{1\over 8 F^6}\bigg\{\left(\frac{s_{24}}{s_{2|3,4}}-{1\over2}\right)
                 \left(\frac{s_{26}}{s_{2|3,6}}-{1\over2}\right)
				 \left(\frac{s_{28}}{s_{2|3,8}}-{1\over2}\right)
                + \left(\frac{s_{97}}{s_{9|7,8}}-{1\over2}\right)
                 \left(\frac{s_{95}}{s_{9|5,8}}-{1\over2}\right)
                 \left(\frac{s_{93}}{s_{9|3,8}}-{1\over2}\right) \nonumber\\
&&\qquad~      +\left({s_{24} \over s_{2|3,4}}-{1\over 2}\right)
				\left({s_{26} \over s_{2|3,6}}-{1\over 2}\right)
			    \left({s_{79}\over s_{9|7,8}}-{1\over 2}\right)
			  +\left({s_{24} \over s_{2|3,4}}-{1\over 2}\right)
			  \left({s_{59} \over s_{9|5,8}}-{1\over 2}\right)
			  \left({s_{79}\over s_{9|7,8}}-{1\over 2}\right)    \nonumber\\
&&\qquad -{1\over 2}
             \left(\frac{s_{24}}{s_{2|3,4}}-{1\over2}\right)
				 \left(\frac{s_{28}}{s_{2|3,8}}-{1\over2}\right)
            -{1\over 2}\left(\frac{s_{26}}{s_{2|3,6}}-{1\over2}\right)
				 \left(\frac{s_{28}}{s_{2|3,8}}-{1\over2}\right)\nonumber\\
&&\qquad -{1\over 2}\left({s_{79}\over s_{9|7,8}}-{1\over 2}\right)
                 \left(\frac{s_{93}}{s_{9|3,8}}-{1\over2}\right)
             -{1\over 2}\left({s_{59} \over s_{9|5,8}}-{1\over 2}\right)
                 \left(\frac{s_{93}}{s_{9|3,8}}-{1\over2}\right)\nonumber\\
&&\qquad -{1\over 2}\left(\frac{s_{26}}{s_{2|3,6}}-{1\over2}\right)
                 \left({s_{79}\over s_{9|7,8}}-{1\over 2}\right)
            -{1\over 2}\left(\frac{s_{24}}{s_{2|3,4}}-{1\over2}\right)
                 \left({s_{59} \over s_{9|5,8}}-{1\over 2}\right)  \bigg\}\nonumber\\
&&+{s_{35} \over 8 F^6 \,s_{3;5}} \bigg\{\left(\frac{s_{26}} {s_{2|3,6}}-{1\over 2}\right)\left(\frac{s_{28}}{s_{2|3,8}}-{1\over 2}\right)
  										+ \left(\frac{s_{79}} {s_{9|7,8}}-{1\over 2}\right)\left(\frac{s_{9|3,5}}{s_{9|3,8}}-{1\over 2} \right)\nonumber\\ &&\qquad \qquad +\left({s_{26}\over s_{2|3,6}}-{1\over 2}\right)\left({ s_{79} \over s_{9|7,8}}-{1\over 2}\right)
                +\left(\frac{s_{28}}{s_{2|3,8}}-{1\over 2}\right)\left(\frac{s_{9|3,5}}{s_{9|3,8}}-{1\over 2} \right)
                 -\frac{s_{28}}{s_{2|3,8}} \,\frac{s_{9|3,5}}{s_{9|3,8}} \bigg\}\nonumber\\
&&+{s_{46} \over 8 F^6 \,s_{4;6}}\bigg\{\left(\frac{s_{2|4,6}} {s_{2|3,6}}-{1\over 2}\right)\left(\frac{s_{28}}{s_{2|3,8}}-{1\over 2}\right)
   									+\left(\frac{s_{79}} {s_{9|7,8}}-{1\over 2}\right)\left(\frac{s_{39}}{s_{9|3,8}}-{1\over 2}\right)\nonumber\\
&&\qquad\qquad  +\left({s_{2|4,6}\over s_{2|3,6}}-{1\over 2}\right)\left({s_{79} \over s_{9|7,8}}-{1\over 2}\right)
                +\left(\frac{s_{28}}{s_{2|3,8}}-{1\over 2}\right)\left(\frac{s_{39}}{s_{9|3,8}}-{1\over 2}\right)
                - \frac{s_{28}}{s_{2|3,8}}\,\frac{s_{39}}{s_{9|3,8}}\bigg\}\nonumber\\
&&+{s_{57} \over 8 F^6 \,s_{5;7}}\bigg\{\left(\frac{s_{24}} {s_{2|3,4}}-{1\over 2}\right)\left(\frac{s_{28}}{s_{2|3,8}}-{1\over 2}\right)
 								       +\left(\frac{s_{9|5,7}} {s_{9|5,8}}-{1\over 2}\right)\left(\frac{s_{39}}{s_{9|3,8}}-{1\over 2} \right)\nonumber\\
&&\qquad\qquad +\left(\frac{s_{24}} {s_{2|3,4}}-{1\over 2}\right)\left(\frac{s_{9|5,7}} {s_{9|5,8}}-{1\over 2}\right)
 			 +\left(\frac{s_{28}}{s_{2|3,8}}-{1\over 2}\right)\left(\frac{s_{39}}{s_{9|3,8}}-{1\over 2} \right)
			 -\frac{s_{28}}{s_{2|3,8}}\frac{s_{39}}{s_{9|3,8}}\bigg\}\nonumber\\
&&+{s_{68} \over 8 F^6 \,s_{6;8}}\bigg\{\left(\frac{s_{24}} {s_{2|3,4}}-{1\over 2}\right)\left(\frac{s_{2|6,8}}{s_{2|3,8}}-{1\over 2} \right)
									   +\left(\frac{s_{59}} {s_{9|5,8}}-{1\over 2}\right)\left(\frac{s_{39}}{s_{9|3,8}}-{1\over 2}\right)\nonumber\\
&&\qquad\qquad    +\left({s_{24}\over s_{2|3,4}}-{1\over 2}\right)\left({s_{59} \over s_{9|5,8}}-{1\over 2}\right)
									   +\left(\frac{s_{2|6,8}}{s_{2|3,8}}-{1\over 2} \right)\left(\frac{s_{39}}{s_{9|3,8}}-{1\over 2}\right)
									   -\frac{s_{2|6,8}}{s_{2|3,8}}\frac{s_{39}}{s_{9|3,8}}\bigg\}\nonumber\\
&&  -\frac{s_{37}}{8 F^6 s_{3; 7}}\left(\frac{s_{28}}{s_{2|3,8}}+\frac{s_{9|3,7}}{s_{9|3,8}}-1\right)
       -\frac{s_{48}}{8 F^6\, s_{4; 8}}\left(\frac{s_{2|4,8}}{s_{2|3,8}} +\frac{s_{39} }{s_{9|3,8}}-1\right)
         \nonumber\\
&&+\frac{s_{35}s_{68}}{8 F^6 \, s_{3;5}\, s_{6;8}}\left(\frac{s_{2|6,8}}{s_{2|3,8}}+\frac{s_{9|3,5}}{s_{9|3,8}}-1\right)\nonumber\\
&&+\frac{s_{35} s_{7|3,5}}{8 F^6\, s_{3;5}\,s_{3;7} }\left(\frac{s_{28} }{s_{2|3,8}}-\frac{s_{89}}{s_{9|3,8}}\right)
  +\frac{s_{46}  s_{37} }{8 F^6\, s_{4;6}\,s_{3;7} }\left(\frac{s_{28} }{s_{2|3,8}}-\frac{s_{89}}{s_{9|3,8}}\right)
  +\frac{s_{57}  s_{3|5,7} }{8 F^6\, s_{5;7}\,s_{3;7} }\left(\frac{s_{28} }{s_{2|3,8}}-\frac{s_{89}}{s_{9|3,8}}\right)\nonumber\\
&&+\frac{s_{46} s_{8|4,6}}{8 F^6\, s_{4;6}\,s_{4;8} }\left(\frac{s_{39}}{s_{9|3,8}}-\frac{s_{23} }{s_{2|3,8}}\right)
  +\frac{s_{57}  s_{48} }{8 F^6\, s_{5;7}\,s_{4;8} }\left(\frac{s_{39}}{s_{9|3,8}}-\frac{s_{23} }{s_{2|3,8}}\right)
  +\frac{s_{68}  s_{4|6,8} }{8 F^6\, s_{6;8}\,s_{4;8} }\left(\frac{s_{39}}{s_{9|3,8}}-\frac{s_{23} }{s_{2|3,8}}\right),\nonumber\\
\eea
where $s_{i|j,k}=2p_i\cdot (p_j+\dots+p_k)$ and $s_{i;j}=(p_i+\dots+p_j)^2$.

\bibliographystyle{JHEP}
\bibliography{Multisoft}

\end{document}